\documentclass{emulateapj}
\usepackage{natbib}
\usepackage{graphicx}
\usepackage{amssymb}
\usepackage{epstopdf}
\usepackage{subfigure}
\usepackage{graphicx}
\usepackage{soul}

\usepackage{color}

\begin{document}


\title{TeV cosmic-ray anisotropy from the magnetic field at the heliospheric boundary}

\author{V. L\'opez-Barquero$^{1}$, S. Xu$^{2}$, P. Desiati$^{3,4}$, A. Lazarian$^{4}$, N. V. Pogorelov$^{5,6}$, H. Yan$^{7,8}$}
\affiliation{
1. Department of Physics, University of Wisconsin, Madison, Wisconsin 53706, USA \\
2. Department of Astronomy, School of Physics, Peking University, Beijing 100871, China\\
3. Wisconsin IceCube Particle Astrophysics Center (WIPAC), University of Wisconsin, Madison, WI 53703, USA\\
4. Department of Astronomy, University of Wisconsin, Madison, WI 53706, USA\\
5. Department of Physics, University of Alabama in Huntsville, Huntsville, AL 35899, USA\\
6. Center for Space Plasma and Aeronomic Research, University of Alabama in Huntsville, 320 Sparkman Dr., Huntsville, AL 35805, USA\\
7. DESY, Platanenallee 6, 15738 Zeuthen, Germany\\
8. Institut fur Physik und Astronomie, Universit{\"a}t Potsdam, 14476 Potsdam-Golm, Germany}


\begin{abstract}

We performed numerical calculations to test the suggestion by~\cite{desiati_lazarian_2013} that the anisotropies of TeV cosmic rays may arise from their interactions with the heliosphere. For this purpose, we used a magnetic field model of the heliosphere and performed direct numerical calculations of particle trajectories. Unlike earlier papers testing the idea, we did not employ time-reversible techniques that are based on Liouville's theorem. We showed numerically that for scattering by the heliosphere the conditions of Liouville's theorem are not satisfied and the adiabatic approximation and time-reversibility of the particle trajectories are not valid. Our results indicate sensitivity to the magnetic structure of the heliospheric magnetic field, and we expect that this will be useful for probing this structure in future research. 
\end{abstract}

\section{Introduction}
\label{sec:intro}

Cosmic rays (henceforth, CRs) with energy below 10$^{18}$ eV have a gyroradius smaller than the galactic disk thickness of about 300 pc, with energy-dependent confinement within the Milky Way. The topics related to the origin, propagation and acceleration of CRs are still debated in spite of the long history of relevant studies (see the excellent textbook by~\cite{longair_2011} and references therein). However, it is generally accepted that most of the galactic CRs are being accelerated by supernova shocks. Some percentage of the CRs can be accelerated by magnetic reconnection~\citep{deg_lazarian_2005}\footnote{Magnetic reconnection becomes fast, i.e., independent of resistivity, in turbulent media~\citep{lv_1999,kowal_2009,kowal_2012} (see also review by~\cite{lazarian_2015}  and references therein). A similar acceleration mechanism that appeals to tearing is discussed in a later publication by~\cite{drake_2006}.}. Spatially, supernovae are correlated to star-forming regions, so
the distribution of CRs is affected by that of their sources, but it is modified by propagation through the galactic magnetic field.
Frequently the magnetic field in the Galaxy is described as composed of a global regular component~\citep[see, e.g.,][]{jansson_farrar_2012a,jansson_farrar_2012b}, large-scale coherent (on the order of 10-100 pc) structures, and the ubiquitous turbulent component (with wide spatial-scale inertial range with amplitude following a Kolmogorov power spectrum).
This is, however, an approximation, with the availability of the modern theory of magneto-hydrodynamic (MHD) turbulence (see~\cite{branden_lazarian_13} for a review) predicting a more sophisticated picture, with compressible and incompressible modes having their own cascades~\citep{gs_1995}, henceforth GS95 \citep{lg_2001,cho_lazarian_2002,cho_lazarian_2003,kowal_2009}. For sub-Alfvenic turbulence, which is typical within quiescent regions of the interstellar medium (ISM), the transition from weak turbulence to strong Alfvenic turbulence takes place~\citep{lv_1999,galtier_2000}. The latter has the Kolmogorov-type spectrum $\sim k^{-5/3}$.
However, this spectrum is strongly anisotropic and therefore the scattering by Alfvenic turbulence injected at large scale is marginal~\citep{chandran_2000,yan_lazarian_2002}, with fast modes identified by~\cite{yan_lazarian_2002} as the major scattering component induced by the galactic turbulent cascade. 

An additional scattering emerges from CR instabilities. Streaming instability has been long considered an important component of CR propagation physics (see~\cite{cesarsky_1980} for a review). Particle streaming was employed in models such as the leaky box model of propagation to explain the high isotropy of observed CRs. In that model, it was assumed that the streaming instability was suppressed in the partially ionized galactic disk and acts to scatter and return CRs as they enter the partially ionized galactic halo. The streaming of particles outside the disk was able to naturally explain the observed dipole anisotropies of the observed CR distribution. This model was later challenged in~\cite{farmer_gold_2004}, who performed  calculations of the streaming instability damping by the ISM turbulence and concluded that the streaming is not expected to take place for the levels of turbulent damping that they adopted. More recently, this conclusion was questioned in~\cite{lazarian_2016} where it was shown that for typical halo conditions scattering instability takes place in the galactic halo. Additional instabilities of CRs (see~\cite{lazarian_beresnyak_2006,yan_lazarian_2011}) can act as additional sources of CR isotropization. In this paper, we assume that the influence of scattering induced by the TeV CR instabilities in the vicinity of the heliosphere is negligible. 

The limitation of the traditional models of CR propagation is not only due to scattering physics. In fact, at scales less than the turbulent injection scale, the particles following magnetic field lines experience super-diffusion with respect to the direction of the mean field~\citep{lazarian_yan_2014}. Such effects can be strongly distorted if the synthetic data cubes are used. Therefore, in what follows, we use only the data cubes obtained by direct MHD numerical simulations.

This paper continues our numerical studies of the origin of CR anisotropies observed at Earth. The first paper,~\cite{barquero_2015} (hereafter referred to as LX16), dealt with the effects of interstellar turbulence on the CR propagation and it did not take into account the strong perturbations induced by the heliosphere. In this paper, on the contrary, we focus our attention on the effects arising from the CR interactions with the heliosphere. The idea that the heliosphere can produce strong scattering on CRs, which could be important for explaining the observed high-energy CR anisotropies, was first suggested in~\cite{desiati_lazarian_2013} (hereafter referred to as DL13). This idea was later tested numerically in~\cite{schwadron_2014} and~\cite{zhang_2014}. The difference between our paper and the earlier studies is that we do not assume that Liouville's theorem and the backtracking of particles is valid. In fact, in this paper, we show that the conditions of Liouville's theorem are not satisfied due to the scattering at the heliospheric boundary. Therefore, we adopt a much more time consuming Monte-Carlo approach with the forward tracking of particles. 

The paper is organized as follows. In Section~\ref{sec:problem}, the problem of observed anisotropy of CRs is formulated along with ways to address it, while in Section~\ref{sec:heliomf} the long tail heliospheric model used in this study is described. Section~\ref{sec:prop} describes the particle integration method used and which CR particles are used in the study. In Section~\ref{sec:lt}, we discuss the validity of applying Liouville's theorem in the context of this work. Results are presented in Section~\ref{sec:res} and discussed in Section~\ref{sec:disc}.  Concluding remarks follow in Section~\ref{sec:concl}.

\section{The problem of anisotropies and corresponding approaches}
\label{sec:problem}

Particle energy roughly determines which spatial scale is the most dominant in shaping the characteristics of their distribution.
%
Galactic CRs in the energy range below about 50 GeV are strongly affected by modulations of the inner heliospheric magnetic field, in correlation with solar cycles (see, e.g.,~\cite{florinski_2013,potgieter_2013,manuel_2014,florinski_2015}). Above 50 GeV, the modulation in the CR energy spectrum is negligible; however, the effects of long-term solar cycles on particle distribution is still observed up to an energy of few hundred GeV~\citep{munakata_2010,kozai_2014}. The gyroradius of 10- to 100-GeV CR particles in the interplanetary magnetic field of $<$ 1 $\mu$G is typically smaller than the size of the termination shock \citep[about 80--90~AU, see][]{pogorelov_2015}. 
This makes it possible for those particles to be spatially redistributed according to the modulating solar wind-induced perturbations on the magnetic field.

At TeV energies, a scale transition occurs. The typical particle gyroradius is larger than the size of the termination shock, therefore the influence of inner heliospheric magnetic fields on the CR distribution is negligible. In fact, solar cycle modulations are subdominant in this energy range. However, TeV galactic CRs coming from the ISM, where the local interstellar magnetic field (LIMF) is $\sim$3 $\mu$G, and propagating into the heliosphere have gyroradius on the order of 100 AU, which is smaller than the estimated thickness of the heliosphere of about 600 AU and shorter than the estimated length of the heliospheric tail of a few thousands AU~\citep{pogorelov_2006,pogorelov_2015,pogorelov_2016}. From this scaling relationship, TeV CRs are expected to be influenced by the heliospheric magnetic field (see~\cite{desiati_lazarian_2013}, hereafter referred to as DL13).
At an energy scale of tens of TeV, the gyroradius of the lightest CR particles starts to exceed the heliosphere's thickness, thus decreasing its influence; nonetheless, these particles will still experience the influence of the perturbation created by the 
heliosphere on the local interstellar medium (LISM) and the effects of the heliotail's length on their propagation.
At higher energies, the arrival directions of the CRs are influenced by their propagation through the interstellar magnetic field (see LX16) and by the distribution of their sources in the Galaxy.

It is, therefore, expected that CRs with energy below several tens of TeV are influenced by the heliosphere to some extent.  The actual degree of such an effect depends on the properties of the heliosphere, such as its size and the magnetic structure, the presence of magnetic perturbations or instabilities at the boundary with the ISM (e.g., at the flanks of the heliosphere), the large-scale perturbation of the LISM due to the heliosphere, 
and the mass composition (or better the rigidity\footnote{Rigidity of a charged particle is a measure of its momentum, and it refers to the fact that a higher momentum particle has a higher resistance to deflection by a magnetic field. It is defined as $R = r_L\,B\,c = E/Ze$, with $r_L$ the particle gyroradius and $B$ the magnetic field. A 1-TeV proton and a 26-TeV iron nucleus have rigidity of 1 TV.}) of the CR particles. If the heliosphere has the effect to redistribute the TeV CR arrival direction distribution, compared to that shaped by interstellar propagation, all those details need to be properly understood and integrated into a comprehensive numerical particle trajectory integration.

From an observational point of view, a statistically significant anisotropy has been observed by a variety of experiments, sensitive to different energy ranges (from tens of GeV to a few PeV), located on or below the Earth's surface in the Northern Hemisphere~\citep{nagashima_1998,hall_1999,amenomori_2005,amenomori_2006,guillian_2007,abdo_2009,aglietta_2009,zhang_2009,munakata_2010,amenomori_2011,dejong_2011,shuwang_2011,bartoli_2015} and in the Southern Hemisphere~\citep{abbasi_2010,abbasi_2011,abbasi_2012,aartsen_2013,aartsen_2015}.

The global anisotropy appears to change with energy in a nontrivial way. From about 100 GeV to tens of TeV, it has an approximately consistent structure at the largest scale, although its measured amplitude increases with energy. Above a few tens of TeV, the observed progressive change in the anisotropy topology may indicate a transition between two processes shaping the particles' arrival distribution at Earth, for instance, the transition from heliospheric-dominated to ISM-dominated influence, which culminates around 100 TeV (as observed in~\cite{aartsen_2015} and discussed in DL13).

However, the change in topology of the CR anisotropy as a function of energy can have different origins as well. In the scenario of particles in homogeneous and isotropic diffusion, the CR density gradient, and therefore the induced spatial anisotropy, has a dipolar shape. The direction of the dipole is expected to point towards the strongest source of the observed CRs, and its amplitude to depend on the diffusion coefficient. At different energies, the strongest contribution to the observations can shift from one source to another, thus changing the orientation of the dipole~\citep{erlykin_2006,blasi_2012,ptuskin_2012,pohl_2013,sveshnikova_2013,savchenko_2015,ahlers_2016}. The difficulty with this scenario is that particle diffusion in the ISM is expected to be anisotropic, i.e., fast along and slow across the magnetic field lines. A misalignment between the CR density gradient and the regular galactic magnetic field prevents pointing to any specific source, and it would suppress the anisotropy amplitude depending on the misalignment angle~\citep{effenberger_2012,kumar_2014,schwadron_2014,mertsch_2015}. Since the ratio of perpendicular to parallel diffusion is likely to depend on energy (depending on the magnetic field geometric configuration), the change in orientation of the anisotropy is also linked to the properties of the interstellar magnetic field itself.

The observed anisotropy cannot be described with a simple dipole component. The actual distribution is a combination of several angular scales~\citep{amenomori_2007,abdo_2008,abbasi_2011,bartoli_2013,abeysekara_2014,aartsen_2015} that can be studied by decomposing it into individual spherical harmonic contributions. This makes it possible to determine the angular power spectrum of the observed arrival distribution. As reported by experimental observations, most of the power is concentrated in the large-scale anisotropy structure, which includes dipole, quadrupole, and octupole. Such contributions are likely affected by the limited field of view of the experiments and also by biases that limit the observation at large scale (see, e.g.,~\cite{ahlers_etal_2016}). About 1\% of the power is distributed across small-scale structures in the arrival direction distribution (where there is no bias due to the field of view).
Small angular scale anisotropy features correspond to regions where CR flux has large gradients in a relatively localized area in the field of view of the observations (on the order of 10$^{\circ}$). Such regions can be stochastically produced by scattering processes of CRs in the ISM magnetic turbulence within the particle mean free path, as discussed in~\cite{giacinti_sigl_2012,ahlers_2014,ahlers_mertsch_2015} and our companion paper LX16. 
Such scattering processes have the effect of decomposing a large-scale particle density gradient into small-scale components. This process constitutes an important contribution to the power spectrum, and it is certainly compatible with observations. However, it is possible to argue that some observed localized regions of TeV CR excess appear to be correlated with features associated with the heliosphere. For instance, one of the localized excess regions observed in the northern equatorial sky appears to be correlated with the direction of the heliospheric tail (see, e.g.,~\cite{amenomori_2007,abdo_2008,bartoli_2013,abeysekara_2014}). CRs observed within this localized region have an energy spectrum that is harder than that in the surrounding areas. It was proposed that reacceleration of CRs by magnetic reconnections in the heliospheric tail may be a possible explanation~\citep{lazarian_desiati_2010,desiati_lazarian_2012}. Other localized regions are spatially correlated with the large angular gradient edge of relative intensity across the whole sky, with a possible link to heliospheric origin (as discussed in DL13).

Magnetic instabilities that dynamically develop at the boundary between the heliosphere and the ISM~\citep{liewer96,zank96,Zank99,Florin05,pogorelov_2006,Borov08,zank09,shaikh_zank_2010,pogorelov_2015} (see Section~\ref{sec:heliomf}) have spatial scales on the order of 10-100 AU, and induce scattering processes on multi-TeV-scale CRs that cross the heliosphere. The possibility that strong resonant scattering processes cause a redistribution of the CR arrival direction distribution was already discussed in DL13. As mentioned, other authors have studied the effects of the heliosphere~\citep{schwadron_2014,zhang_2014} or, in general, of astrospheres~\citep{scherer_2016} on the distribution of TeV CRs
%

The heliospheric model used in the present work makes use of ideal MHD treatment of ions and of a kinetic multi-fluid description of neutral interstellar atoms penetrating into the heliosphere~(\cite{pogorelov_2013}). This model incorporates the heliospheric tail up to a distance of approximately 4000 AU, which does not cover the maximum possible extension (see Section~\ref{sec:heliomf}). 
In this study, the possibility that resonant scattering processes may have a strong effect on redistributing TeV CR arrival direction distribution is critically discussed.
The relevant points of the present study are to dispute whether Liouville's theorem can actually be used as a tool to determine the particle trajectories affected by the heliospheric magnetic field and whether the heliosphere itself imprints a strong effect on the cosmic particles crossing it.
If magnetic fields change significantly within gyroradius spatial scale, the geometry of particle trajectories may be highly sensitive to the actual initial conditions; i.e., they may have a chaotic nature. In such a case, application of Liouville's theorem is not warranted and particle distribution may follow a different scaling. In general, application of Liouville's theorem must to be investigated case by case. 
In what follows, we use direct numerical simulations of CR propagation using the numerical results of heliospheric simulations and without any use of  Liouville's theorem.

\section{Cosmic-ray propagation in the heliosphere}
\label{sec:prop}

In this section, the description of the heliospheric magnetic field model used in the present study is laid out, then the strategy and method used to numerically integrate the particle trajectories through the heliosphere is described.

\subsection{Heliosphere magnetic field model}
\label{sec:heliomf}

The heliosphere is formed when the solar wind (SW) collides with the local interstellar medium (LISM).
In an ideal magnetohydrodynamic formulation of the problem, the SW--LISM interaction necessarily creates a tangential discontinuity that separates the
plasmas originating at these two sources. This discontinuity is called the heliopause (HP).
The HP extends thousands of astronomical units (AU) from the Sun. As any tangential discontinuity, the HP is subject to
hydrodynamic instabilities, e.g., the Kelvin--Helmholtz (KH) instability \citep[see, e.g.,][]{Belov,Chalov,Ruderman,Ruderman10}.
Moreover, the HP nose is subject to Rayleigh--Taylor (RT) instability. The role of gravity in this case is played by the momentum-exchange
terms in the MHD equations describing the plasma flow in the presence of charge exchange \citep{liewer96,Zank99}. Linear analysis \citep{Avinash} of the
RT instability performed in an idealized formulation showed that perturbations grow unconditionally while they are small.
Nonlinear, numerical investigations of the RT instability have been performed by \citet{Florin05} and \citet{Borov08} in an axially-symmetric case,
and by \citet{Borov14} in a realistically three-dimensional formulation. It was demonstrated in the latter paper that the heliospheric magnetic
field (HMF) can damp the RT instability. However, the HMF becomes rather small occasionally in the course of solar cycle, so the instability results
in a substantial mixing of LISM and SW plasma at the nose of the heliopause. Interestingly, \citet{Borov14} shows that RT instability
may reveal itself also at the HP flanks, but it is caused by charge exchange with secondary neutral atoms born inside the heliosphere.
As a result, the HP surface bounding the heliotail is subject to the mixture of KH and RT instabilities.

Since the KH instability is of a convective type, with the perturbation amplitude growing as a function of distance along the HP, plasma mixing as well as diffusion
processes would eventually destroy the HP. However, it appears that charge exchange is a dominant process that results
in continuous elimination of the hotter SW ions with cooler ions possessing properties of the LISM H atoms. As shown by \citet{Izmod03} in an axially-symmetric case
and \citet{Pogo15} in 3D, this makes the SW flow superfast magnetosonic (its Mach number calculated using the fast magnetosonic speed is greater than 1)
at about 4,000~AU. In the investigation presented in this paper, we use one of the heliotail models described in \citet{Pogo15}.
This model is based on a self-consistent solution of the ideal MHD equations, with appropriate source terms due to charge exchange between ions and neutrals,
to describe the flow of plasma and the Boltzmann equation to describe the transport of neutral atoms.
To avoid issues related to the heliospheric current sheet, this model assumes a unipolar distribution of the HMF inside the heliosphere.
It shows initial collimation of the SW plasma inside the Parker spiral field lines bent tailward by the flow, as predicted by \citet{Yu}.
The spiral field being kink-instable \citep{Roberts,Pogo15}, the reason for such collimation disappears at about 800~AU from the Sun.
In contrast with \citet{Yu} and multi-fluid simulations of \citet{Opher15}, no separation of the HP into two lobes occurs in \citet{Pogo15}. This is because
multi-fluid models substantially depress charge exchange across the lobe separation region. Such entirely hydrodynamic artifacts are impossible
if atoms are treated kinetically. We assume the following properties of the LISM: temperature $T_\infty=6300$~K, velocity $V_\infty=23.2$ km/s, proton density
$n_\infty=0.082$ cm$^{-3}$, H atom density $n_{H\infty}=0.172$ cm$^{-3}$, and magnetic field strength $B_\infty=3\ \mu\mathrm{G}$.
The LISM flow comes from the direction $(\lambda, \beta)=(79^\circ, -5^\circ)$, while the $\mathbf{B}_\infty$ vector arises from
$(\lambda, \beta)=(225^\circ, 44^\circ)$ in the ecliptic coordinate system \citep{Zank13}.
The SW is assumed to be spherically symmetric with the following properties at 1~AU: plasma density $n_\mathrm{SW} = 7.4$ cm$^{-3}$, temperature $T_\mathrm{SW} = 51100$~K,
radial velocity $V_\mathrm{SW} = 450\ \mathrm{km}\,\mathrm{s}$, and radial magnetic field component $B_R = 37.5\ \mu\mathrm{G}$. The HMF is assumed to be
Parker's at 1~AU.

\subsection{Particle trajectory integration}
\label{ssec:crprop}
%
\begin{figure}[t!]
\includegraphics[width=0.89\columnwidth]{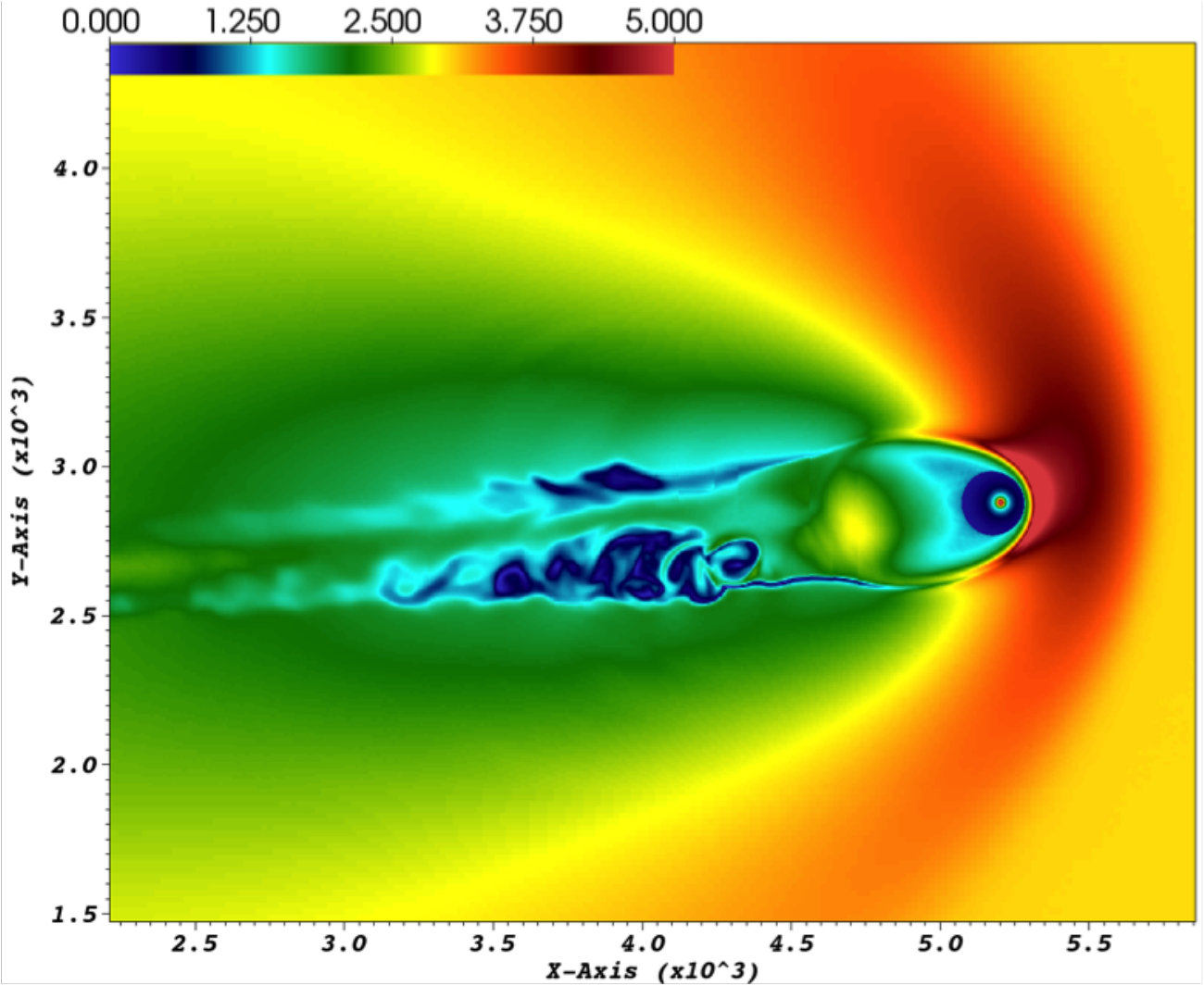}
\includegraphics[width=\columnwidth]{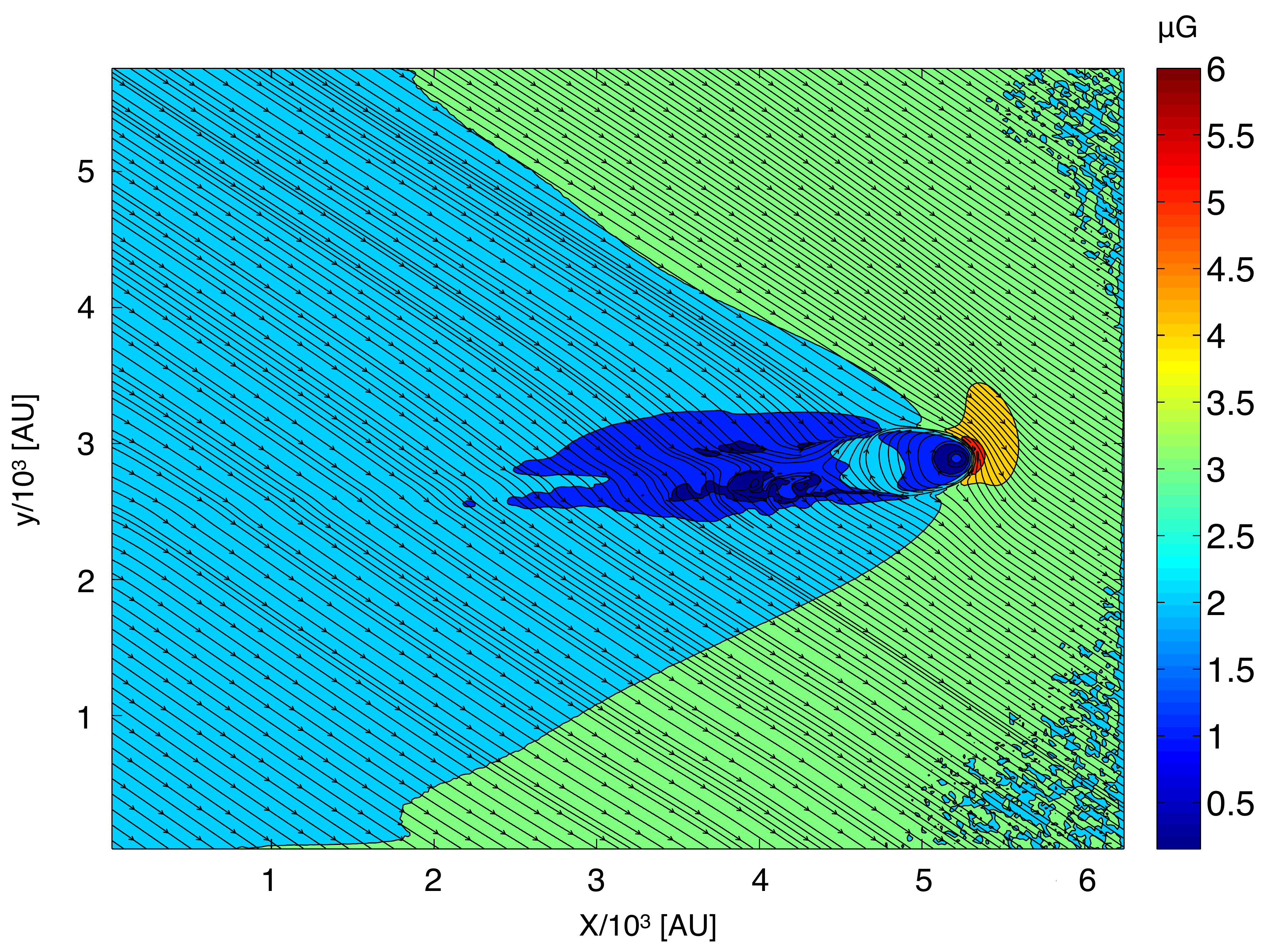}
\caption{Meridional projection of the heliospheric magnetic field model described in Section~\ref{sec:heliomf}. At the top, the figure shows high-resolution visual details of the magnetic fields (color code in units of $\mu$G) with the axes in units of AU. At the bottom, the figure shows the magnetic field lines (coarse color code in units of $\mu$G) with the axes in units of AU. The simulation box is 320 $\times$ 280 $\times$ 280 grid points, corresponding to 6400 AU $\times$ 5600 AU $\times$ 5600 AU (20 AU/grid point).}
\label{fig:heliofield}
\end{figure}
Similar to the work in LX16, this study is performed by integrating particle trajectories in the magnetic field described in Section~\ref{sec:heliomf}, using the set of 6-dimensional ordinary differential equations
\begin{eqnarray}
\frac{d\vec{p}}{dt} & = & q \left( \vec{u}\times\vec{B} \right)\\
\frac{d\vec{r}}{dt}  & = & \vec{u},
\label{eq:motion}
\end{eqnarray}
describing the Lorentz force with $\vec{u}$ the particle velocity, $\vec{r}$ its position vector and $\vec{p}$ the momentum. For $\vec{B}$, we use one steady-state realization of the heliospheric magnetic field described in section~\ref{sec:heliomf}.
As in LX16, the equations are integrated using the Bulirsch-Stoer integration method with adaptive time step. At each integration step, the magnetic field is interpolated using a 3D cubic spline, and integration is stopped when particles cross the border of the simulation box.
%
%
The choice of one specific realization of the magnetic field is justified by the fact that particle velocity is much larger than the plasma Alfv\'en velocity; thus, induced electric fields can be neglected. 


Figure~\ref{fig:heliofield} shows the meridional projection of the heliospheric model snapshot (described in section~\ref{sec:heliomf}) used in this study. At the top, the figure shows high-resolution visual details of the magnetic fields (color code in units of $\mu$G) with the axes in units of AU. At the bottom, the figure shows the magnetic field lines (coarse color code in units of $\mu$G). 
%
The initial simulation box is 320 $\times$ 280 $\times$ 280 simulation grid points (the longer dimension to allow the inclusion of the heliospheric tail) where each grid point corresponds to 20 AU. This is equivalent to 6400 AU $\times$ 5600 AU $\times$ 5600 AU.
The model has a varying resolution, depending on the region within the heliosphere, with the highest resolution in the region around the Sun, where it is 1 AU (0.05 simulation grid points). The model includes magnetic instabilities, on spatial scales on the order of 10-100 AU, that grow on the flanks at the boundary with the ISM, and a tail with length of approximately 4000 AU. In the model, the uniform interstellar magnetic field outside of the heliospheric boundary has intensity of about 3 $\mu$G. For the study presented in this work, the original simulation box is extended in such a way that a sphere centered at Earth and with radius 6000 AU (300 simulation grid points) is inside it. In the extension of the simulation box, a uniform magnetic field with intensity 3 $\mu$G and with the same direction as in the simulation is assumed.


In order to calculate the distribution of CRs at Earth after their propagation across the interstellar and heliospheric magnetic field, there are two possible methods that can be followed. One is to integrate a large number of particle trajectories (typically isotropically distributed) initiated on Earth and {\it back-propagate} (or {\it back-track}) them to  outer space. The other method is to integrate particle trajectories initiated at a large distance from Earth and {\it forward-propagate} (or {\it forward-track} them, meaning directly propagate them according to the arrow of time) towards Earth. The back-propagation technique is based on the validity of Liouville's theorem, which states that particle density in phase space is conserved along the particle trajectories 
if conditions such as no collisional scattering or no resonant collisionless scattering are satisfied (see section~\ref{sec:lt}). If the theorem conditions are valid, then particle trajectories can be time reversed. This method is very efficient, because it entails the integration of particle trajectories from the target back into outer space. Nevertheless, there is no particle loss, and only portions of space that are directly {\it magnetically connected} to the target location have non-zero particle density population. Therefore, it is necessary to impose an initial anisotropy as a directional dependent weight in order to calculate the particle distribution at the target position. Such a weight function breaks the isotropy initially constructed and provides the anisotropy distribution transmitted back to the target position from the magnetically connected remote regions of space. By construction, this technique does not take into account the generation of anisotropy from particle escape during their propagation.

In the presence of turbulence or instabilities, magnetic fields can vary in spatial scales that are shorter than the particle gyroradius. This breaks adiabaticity and effectively induces collisional processes that may invalidate the application of Liouville's theorem. The validity of Liouville's theorem is extensively discussed in LX16. In Section~\ref{sec:lt} of this paper, it is argued that the theorem does not have the grounds to be applicable for this particular study of heliospheric effects, and therefore it cannot be utilized.
%
In this case, therefore, the forward-propagation method is used. 
Such a technique is implicitly inefficient because only a small fraction of the injected particles will make it to or near the target. As discussed in~\cite{rettig_pohl_2015} as well, this method naturally accounts for the role that CR escape has in shaping the anisotropy. There is no need to assume a global anisotropy at large distance (injected, for instance, by CRs diffusively propagated away from a source) to obtain anisotropy at Earth, since particles naturally stream along interstellar magnetic field lines and undergo scattering processes in magnetic turbulence or instabilities.

In this study, it is argued that scattering at the boundary between the heliosphere and the ISM breaks the particle trajectory reversibility in that those particles that escape without reaching Earth cannot be represented in a back-propagation calculation method. Yet, those trajectory configurations occur and contribute to the overall shape of CR arrival direction distribution.
In addition, 
the distribution of CRs at Earth is reshaped by the heliospheric instabilities in a stochastic manner and the exact individual features produced by this phenomenon may not be predicted; therefore, a statistical approach is used, for instance by calculating the angular power spectrum of the arrival distribution.

The inefficiency intrinsic to the forward-propagating methods is compensated by starting particle trajectory integration only from those areas in space where they have a significantly higher chance to reach the neighborhood of Earth, and by assuming a ``larger size" of Earth to record the trajectories that arrive at the final target. 
Although this last assumption may lead to approximate results, it is sufficient to unveil the role that the heliosphere has on the propagation of TeV CRs independent of the propagation history in the ISM. The actual prediction of the anisotropy details most probably results from fine-tuning of several effects, and it is not addressed in this work.
%

\subsection{Cosmic-Ray Composition}
\label{ssec:crcomp}

\begin{figure}[t!]
\begin{center}
\includegraphics[width=\columnwidth]{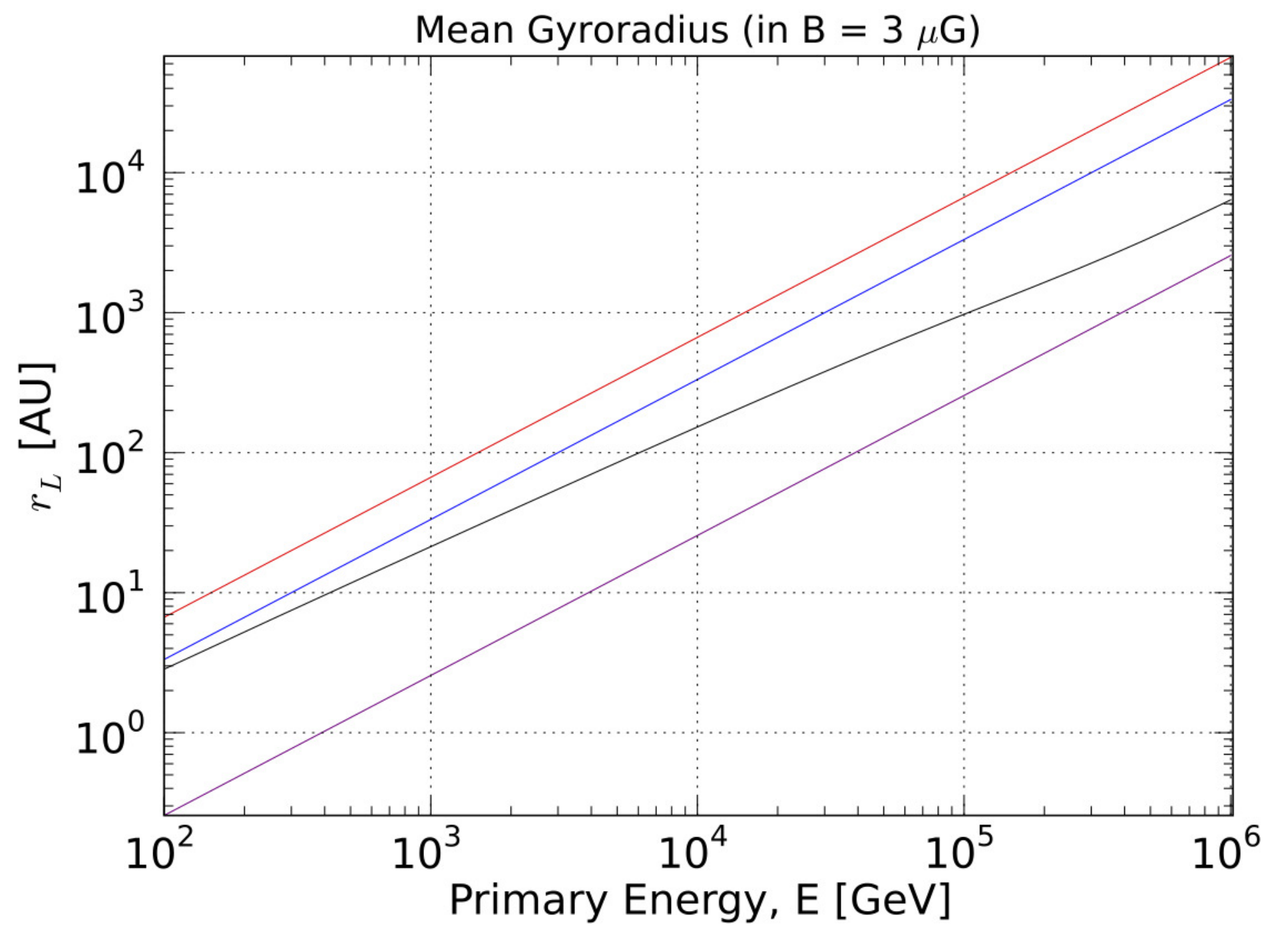}
\end{center}
\caption{Cosmic-ray maximum gyroradius (or Larmor radius) $r_L$ in a 3 $\mu$G magnetic field as a function of particle energy averaged over the observed mass composition (from~\cite{gaisser_2013}) (black line). This is compared to that of protons (red line), of helium (in blue), and of iron nuclei (in purple). Note that due to the mass composition of cosmic rays the average gyroradius is smaller than that for pure protons. This difference becomes important for energies in excess of about 1 TeV.}
\label{fig:gyroradius}
\end{figure}

An important aspect of this study is taking into account that CRs are not dominated by protons only, which is particularly true at energies in excess of 1 TeV. As shown in~\cite{gaisser_2013} and references therein, the abundance of helium nuclei is already comparable to that of protons at the TeV energy range, and it starts to dominate at 10 TeV. Heavier particles become increasingly more important at higher energies as well. Figure~\ref{fig:gyroradius} shows that the maximum CR particle gyroradius $r_L$, averaged over the CR composition, is smaller than that of only protons the higher the contribution from heavier nuclei.
Therefore, the relevant quantity is not the CR particle energy but their rigidity $R = r_L\,B\,c$. Particles with same rigidity have the same gyroradius $r_L$ in a given magnetic field $B$. Or, equivalently, an iron nucleus of energy E has a gyroradius that is 26 times smaller than that of a proton with the same energy and in the same magnetic field B. In the energy range of 1-10 TeV, galactic CRs are approximately composed of a mix of protons, helium, and heavier atomic nuclei~\citep{gaisser_2013}. CR composition is an important ingredient in the understanding of anisotropy, especially since heavier particles, i.e., with small rigidity, may have a non-negligible contribution even at relatively high particle energy.

\subsection{Particle data sets}
\label{ssec:sets}
\begin{figure}[t!]
\begin{center}
\hspace*{-0.85cm}
\includegraphics[width=1.2\columnwidth]{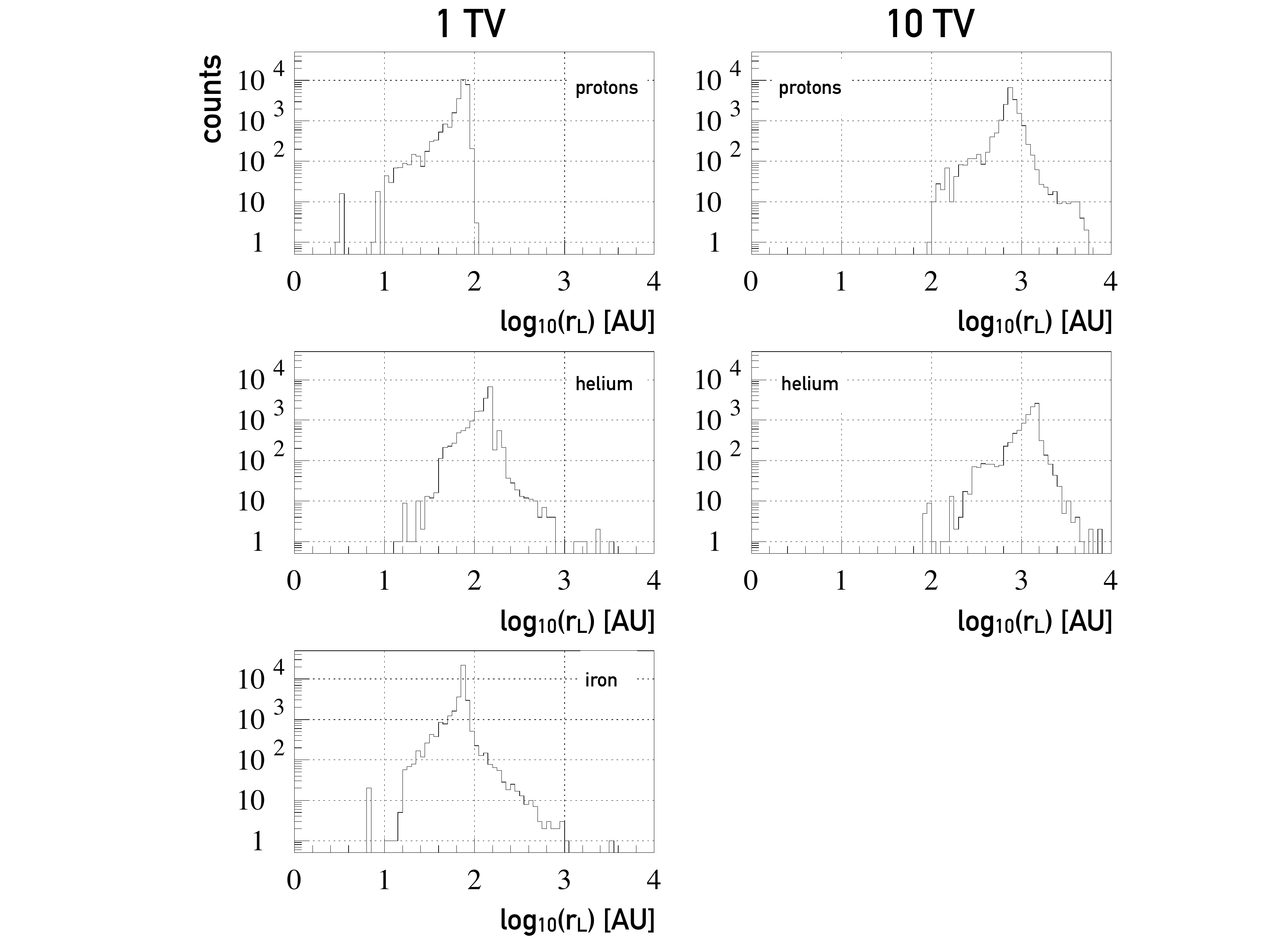}
\end{center}
\caption{Distributions of instantaneous gyroradii $r_L$ (in units of AU) of the particles from sets of Table~\ref{tab:sets} calculated along their trajectories. Note the wide range of variabilities of $r_L$ due to the changes in magnetic field and pitch angle as particles propagate through the heliosphere.}
\label{fig:rl}
\end{figure}
%

\begin{figure}[t!]
\includegraphics[width=\columnwidth]{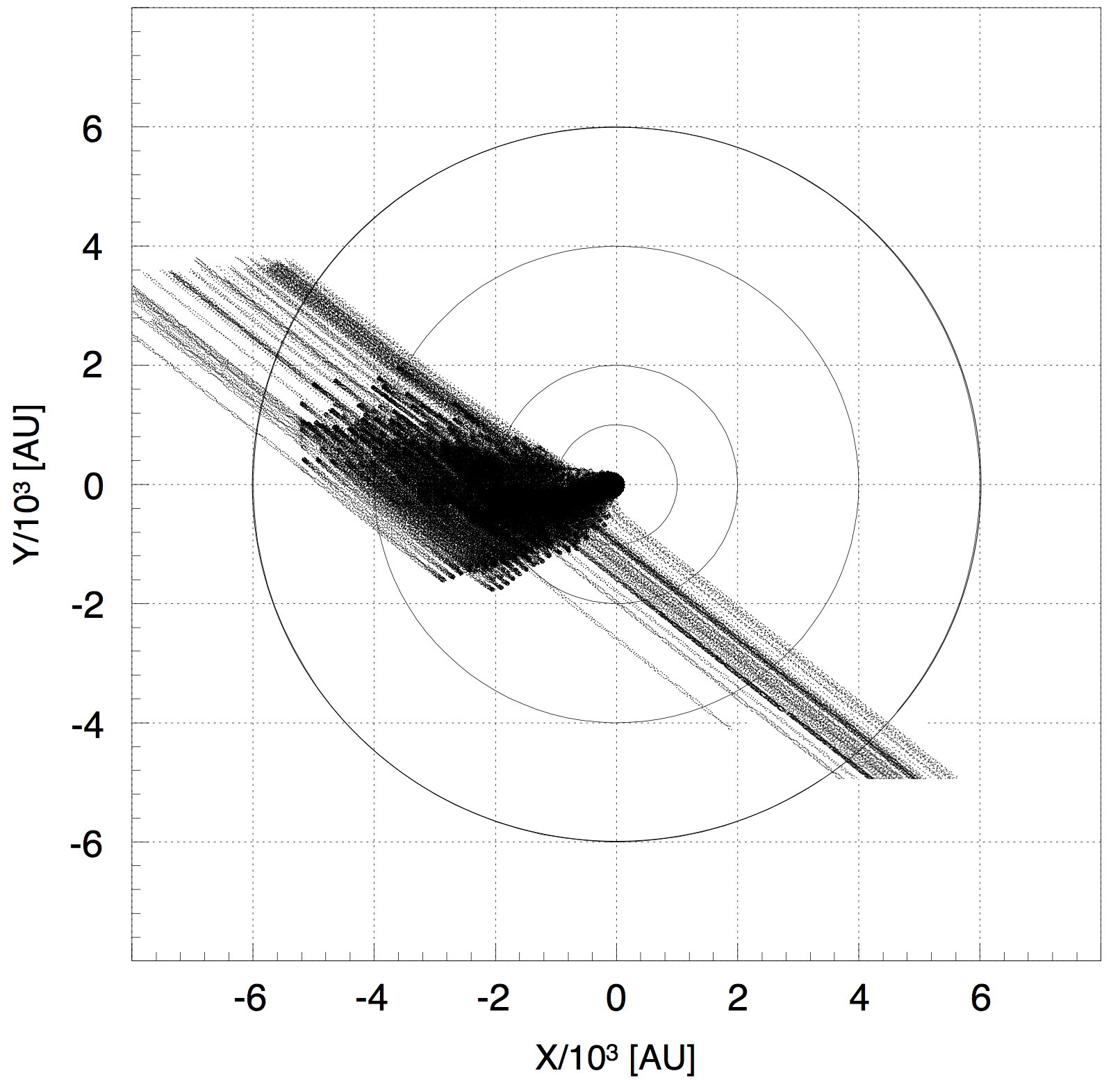}
\caption{Integrated trajectories of protons with energy of 1 TeV, starting from Earth with initial uniform direction distribution, calculated with the heliospheric magnetic field of Figure~\ref{fig:heliofield}. The figure illustrates the complex structure of over 100 trajectories passing through the heliosphere and ultimately streaming along the uniform interstellar magnetic field. The regions where the trajectories cross the injection sphere of radius 6000 AU are used to identify where to forward-propagate cosmic-ray particles (see text). Note that on the interstellar-wind downstream direction (i.e., in the upper left corner of the figure), particles are more spread out in space as an effect of the elongated heliospheric tail, compared to those in the upstream direction (i.e., in the lower right corner of the figure).
}
\label{fig:traj}
\end{figure}

In this study, trajectories for three types of particles are integrated, as shown in Table~\ref{tab:sets}: protons, helium, and iron nuclei. Since the energy range of interest is approximately 1-10 TeV, two rigidity ranges are used here: 1 TV (for all three particle types) and 10 TV (for protons and helium nuclei). For each set, 1$\times$10$^6$ particle trajectories are integrated with initial position on a sphere (labeled as {\it injection sphere}) of radius 6000 AU (300 simulation grid points) centered on Earth and with uniform direction distribution towards the inner volume of the sphere. The initial positions correspond to regions where particles streaming along the LIMF have a higher chance to reach the Earth's neighborhood.
To account for scattering processes, the injection regions, on the interstellar wind upstream and downstream directions of the LIMF, were initially identified by back-propagating particle trajectories from Earth across the heliospheric magnetic field (as shown in Figure~\ref{fig:traj}). Once these main zones were identified, their extension was then expanded to account for the chance that particles initiated further away may reach Earth and yet maintain a manageable efficiency level.
The regions where forward-propagated particle trajectories start are identified with a 30$^{\circ}\times$30$^{\circ}$ zone on the interstellar wind upstream side of the heliosphere (i.e., on the lower right side of Figure~\ref{fig:traj}) and with a 60$^{\circ}\times$60$^{\circ}$ zone on the interstellar wind downstream side of the heliosphere (i.e., on the upper left side of Figure~\ref{fig:traj}). For each set in Table~\ref{tab:sets}, out of 1$\times$10$^6$ particles initiated in both regions, approximately 1$\times$10$^4$ reach the vicinity of Earth (i.e., cross a sphere centered on Earth, labeled as {\it target sphere}, of radius 200 AU (10 simulation grid points)).
%
\begin{table}[!t]
\caption{Physics parameters of simulation sets}
\centering
\begin{tabular}{ lllllll }
\hline
Set	& Particle	& $E_p$		& $\langle r_L\rangle$	& Injected		& Recorded \\
 	&  		&			& 					& particles 	& particles \\
\hline
1	&Protons	& 1 TeV	        & 70.23 $\pm$ 14.63 AU	& $1\times 10^6$	& 8758 \\
2	&Helium	& 4 TeV		& 121.7 $\pm$ 54.5 AU	& $1\times 10^6$	& 10416 \\
3	&Iron     	&30 TeV		& 72.89 $\pm$ 31.50 AU	& $1\times 10^6$	& 6065 \\
\hline
4	&Protons	& 10 TeV		& 764.3 $\pm$ 252.2 AU	& $1\times 10^6$	& 9789 \\
5	&Helium	& 40 TeV		& 1235. $\pm$ 383. AU		& $1\times 10^6$	& 8655 \\
\hline
	&		&			&				& $5\times 10^6$	& 43683 \\
\hline
\end{tabular}
\label{tab:sets}
\end{table}

The integrated trajectories from the sets of Table~\ref{tab:sets} are cumulated in the two rigidity range groups according to the relative mass composition from~\cite{gaisser_2013}: the 1 TV and the 10 TV rigidity scales. Table~\ref{tab:sets} shows the mean instantaneous gyroradius of the numerically integrated trajectories in the heliospheric magnetic field, and the corresponding RMS (numerical values taken from the distributions in Figure~\ref{fig:rl}). The low-rigidity particle group corresponds to $\langle r_L\rangle \sim$ 70-100 AU (i.e., smaller than the size of the heliosphere). For a large fraction of particles, $r_L$ is the same order of magnitude as the magnetic instabilities on the heliospheric boundary. Note the wide distributions of $r_L$ in Figure~\ref{fig:rl} are due to the changes in magnetic field and pitch angle as particles propagate through the heliosphere.
The high-rigidity group corresponds to $\langle r_L\rangle \sim$ 700-1000 AU (i.e., just larger than the heliosphere thickness but smaller than the predicted heliospheric tail length). As discussed in DL13, the rigidity scale of 10 TV is when the heliospheric effect on the particle distribution starts to become subdominant, compared to that from the ISM. This rigidity scale corresponds to CR particles in the range of 10-300 TeV, depending on the mass.


%
%
\section{The validity of Liouville's theorem}
\label{sec:lt}

As discussed in Section~\ref{ssec:crprop}, there are two possible ways to obtain the anisotropy at Earth. 
One involves the application of Liouville's theorem to link the distribution of the particles at some distance in the ISM to the arrival distribution at Earth. The other is to directly propagate the particles from outer space and record the particles' positions at Earth.

In LX16, the theoretical framework for the application of Liouville's theorem in the study of CR arrival direction was provided. The theorem states that the particle density in the neighborhood of a given system in phase space is constant if restrictions are imposed on the system~\citep{goldstein_2002}. 
We obtained the equation: 
\begin{equation}
\label{vlasov}
\frac{\partial \rho }{\partial t} + \vec{v}  \cdot \vec\nabla ( \rho ) + \vec{F}  \cdot  \vec\nabla_{p}( \rho) = \frac{d\rho}{dt} = 0,
\end{equation}
which is precisely the expression for Liouville's theorem~\citep{goldstein_2002, bradt_2008b}, where $\vec\nabla_{p}$ is the {\tt del} operator in momentum space, $\rho$ the distribution function, and $\vec F$ the applied external force.

The most relevant conditions for its application, as shown in LX16, are that the number of particles is conserved and that the forces acting on the particles are p-divergence free. This last restriction tells us that the forces have to be conservative and differentiable. Collision processes evidently violate this condition.
%
In addition, Liouville's theorem can be considered in the context of conservation of information. Each time that a collision event occurs, it violates the connectedness, and information is lost. Therefore, particle trajectories cannot be time reversed.
%

The derivation provided in LX16 for Eq.~$\ref{vlasov}$ is for a pure magnetic force, but in fact, when calculating particles' trajectories in a magnetic field subject to perturbations and instabilities, a variety of factors come into play. 
The most significant effect is when particles encounter a region where the magnetic field varies abruptly, i.e., the scale of variation of the magnetic field is shorter than the gyroradius of the particle. In this scenario, the trajectory does not have time to adjust smoothly to this change, and the scattering process can be effectively considered a collision. For such cases, the right-hand side of Eq. $\ref{vlasov}$ can be modified by the addition of a term,
\begin{math} {\left [ \frac{\partial\rho }{\partial t} \right ]}_c \end{math}, which takes into account collisions of various origins that are differentiated by their exact functional form, given the fact that they will cause a nonzero time rate of change in the distribution function~\citep{baumjohann_1996}. Under these conditions, Liouville's theorem can't be applied.

To test the abruptness in particle trajectories, it is possible to calculate how the density in phase space is modified by scattering processes, i.e., how adiabatic the change is.
%
In the presence of collisions, the magnetic moment of the gyrating particles changes. Therefore, to check for the adiabaticity of the trajectories, we can calculate the magnetic moment for each particle at each time step and find out if, statistically, it truly behaves as an adiabatic invariant.
The relativistic magnetic moment (also called {\it first} adiabatic invariant) is given by:
\begin{equation}
\mu = \frac{{p_{\bot}}^{2}}{2 m |\vec{B}|},
\label{eq:magmom}
\end{equation}
where $p_{\bot}$ is the momentum perpendicular to the magnetic field $\vec{B}$ and $m$ the particle mass. This quantity, relating magnetic field and perpendicular momentum of the particle, is conserved if the field gradients are small within distances comparable to the particle gyroradius. Note that no assumption about the conservation of magnetic moment is used in the numerical integration calculation.

\begin{figure}[t!]
\includegraphics[width=\columnwidth]{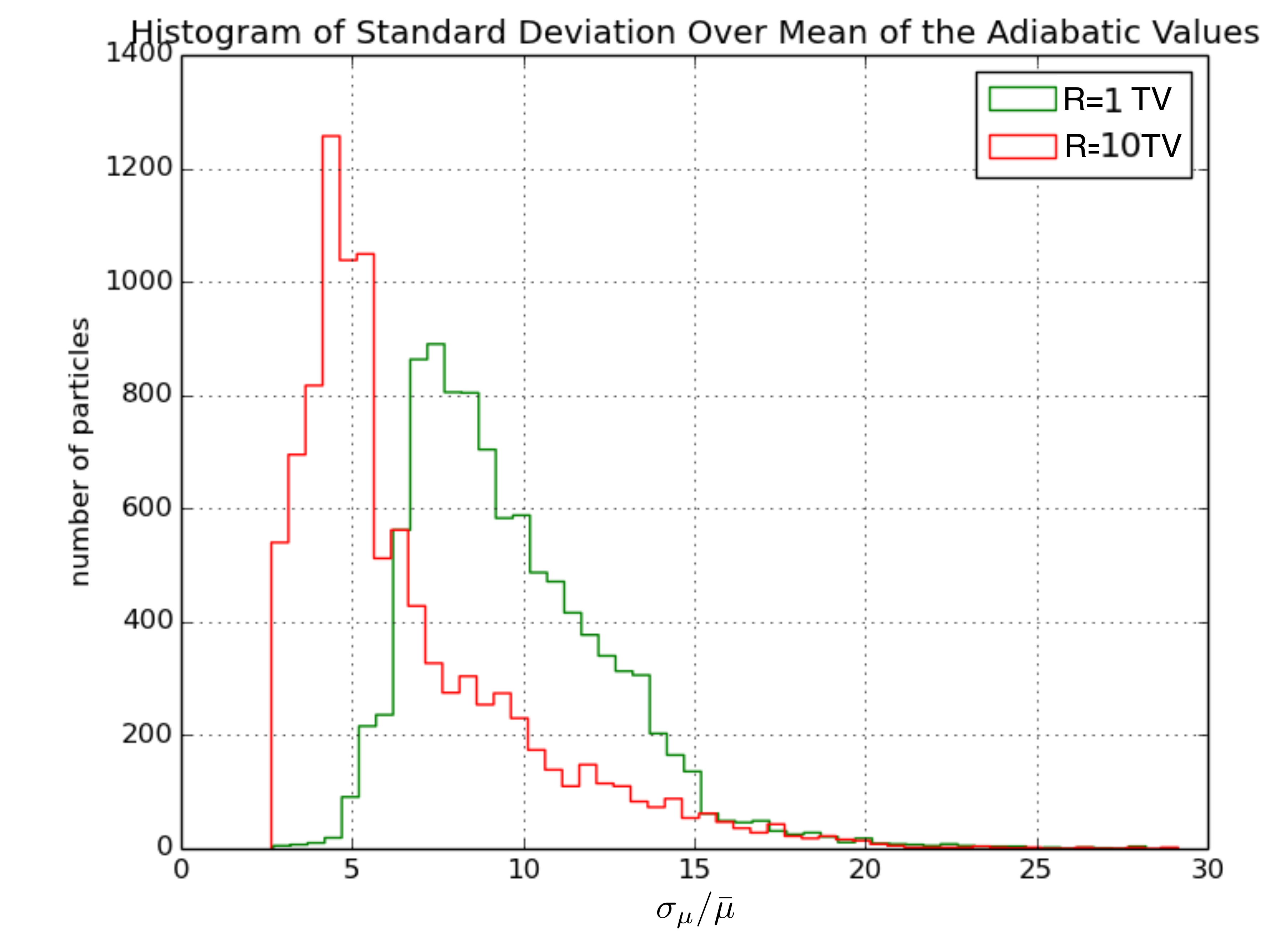}\\
\caption{Histogram of standard deviation of magnetic moment $\sigma_{\mu}$ over mean magnetic moment $\bar{\mu}$ for the two rigidity data sets of Table~\ref{tab:sets}. The red histogram corresponds to the R = 1 TV (p, He, Fe) mixed composition set, and the green histogram to the R = 10 TV (p, He) mixed composition. The magnetic moment is calculated for each particle at all time steps. The mean value and the standard deviation are for the total trajectory.}
\label{fig:promap}
\end{figure}

To perform a statistical test on the first adiabatic invariant, we integrated trajectories from two different data sets, each with $1 \times 10^4$ particles, initiated at Earth and back-propagated to outer space. One set corresponding to 1 TeV and the other one to 10 TeV protons.
Using these specific sets, the magnetic moment in Eq.~\ref{eq:magmom} was calculated at each integration time step and plotted in a histogram. The mean value $\bar{\mu}$ of the magnetic moment of the particle ensemble from each data set and the corresponding standard deviation $\sigma_{\mu}$ were calculated. Figure~\ref{fig:promap} shows the ratio $\sigma_{\mu}$/$\bar{\mu}$ for the two sets.
A distribution with $\sigma_{\mu}$/$\bar{\mu}$ = 0 indicates that the conservation of the magnetic moment is perfect. In this case, particles can mirror back and forth between magnetic bottles, which maintain magnetic moment conserved.
However, a distribution peaked at a value much larger than one means that the particles suffer strong variations in their trajectories and collision-like interactions happened. 
It is not simple mirroring for most particles, but effectively resonant scattering processes at the heliospheric boundary where particles propagate across magnetic field lines (see, e.g.,~\cite{desiati_zweibel_2014}) with a stochastically redistributed pitch angle.
The distributions obtained for these two sets show peaks at around 5 and 7, with strong skewness to the right, or towards higher values of
$\sigma_{\mu}$/$\bar{\mu}$. This indicates that the magnetic moment fluctuated strongly, that severe changes have occurred, and collision-like interactions happened to the particles under consideration.
A stronger deviation in the conservation of magnetic moment is observed with the low-rigidity set than with the high-rigidity set, because of the stronger scattering effects at lower rigidity.
Note that in LX16 the same trajectory integration code was used to calculate particle trajectories in compressible MHD turbulence, in the rigidity range between 750 TV and 30 PV. In that case, it was found that the ensemble average was $\langle \sigma_{\mu}$/$\bar{\mu}\rangle \lesssim$ 1, i.e., significantly smaller than in the present case. The accuracy of the numerical trajectory integration is at the same level as that in LX16. The adaptive time step algorithms constrain both spatial and momentum coordinates to the same relative error level, thus limiting accuracy in spatial coordinates to a level that is $\ll r_L$, even after several tens of thousands of gyrations. Therefore, the strong nonconservation of the ensemble-average magnetic moment is not caused by lack of numerical accuracy but rather by the characteristics of the magnetic field used in this study. The scattering processes with the heliospheric magnetic instabilities determine the global statistical properties of the particles. The value of $\langle \sigma_{\mu}$/$\bar{\mu}\rangle$ in Figure~\ref{fig:promap} is smaller in the higher rigidity set because scattering is less effective in redistributing particles with a gyroradius significantly larger than the spatial size of the instabilities.
%


To conclude, interactions with the heliospheric magnetic field model used in this study result in dramatic changes in the distribution of particle trajectories. The original directional information carried by the particles is lost in these collision-like events. Trajectories diverge due to the magnetic field lines geometry in the regions of magnetic instabilities at the boundary between the heliosphere and the ISM.
Thus, based on the above considerations, it is not possible to assume that Liouville's theorem is applicable in this case.
%
%

For that reason, in this work, the forward propagation method is used.
With this method, 
anisotropy arises naturally from particle propagation and the interaction with the heliosphere. One important factor to take into account is that since there is a violation of the conditions of Liouville's theorem, we cannot make causal links to or rely in any way on the reversibility of the trajectories; consequently, this constrains the possibility of connecting a specific position in the ISM and the arrival direction at Earth. Therefore, it is only possible to determine to what degree the incoming distribution from outer space is distorted due to the features of the heliosphere but not possible to draw direct correlations between the incoming specific directions and the ones observed at Earth. In our case, we will show how the heliosphere acts on and distorts this distribution and how anisotropies arise.  








%
\section{Results}
\label{sec:res}

This section shows the results obtained with the numerical calculation data sets described in Section~\ref{ssec:crprop}.

\subsection{Sky Maps of Arrival Direction Distribution}
\label{sec:skymaps}

\begin{figure}[t!]
\includegraphics[width=\columnwidth]{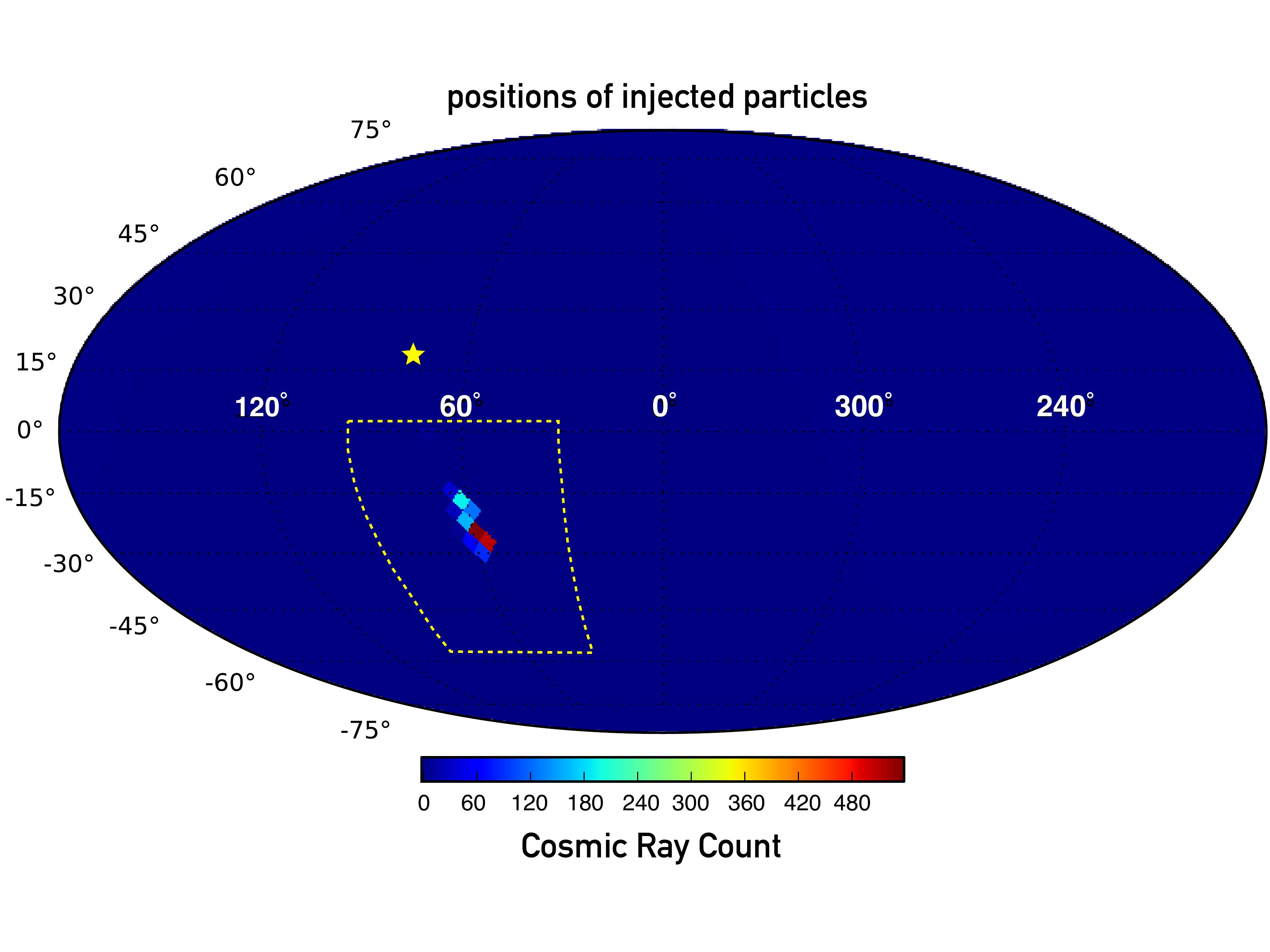}
\includegraphics[width=\columnwidth]{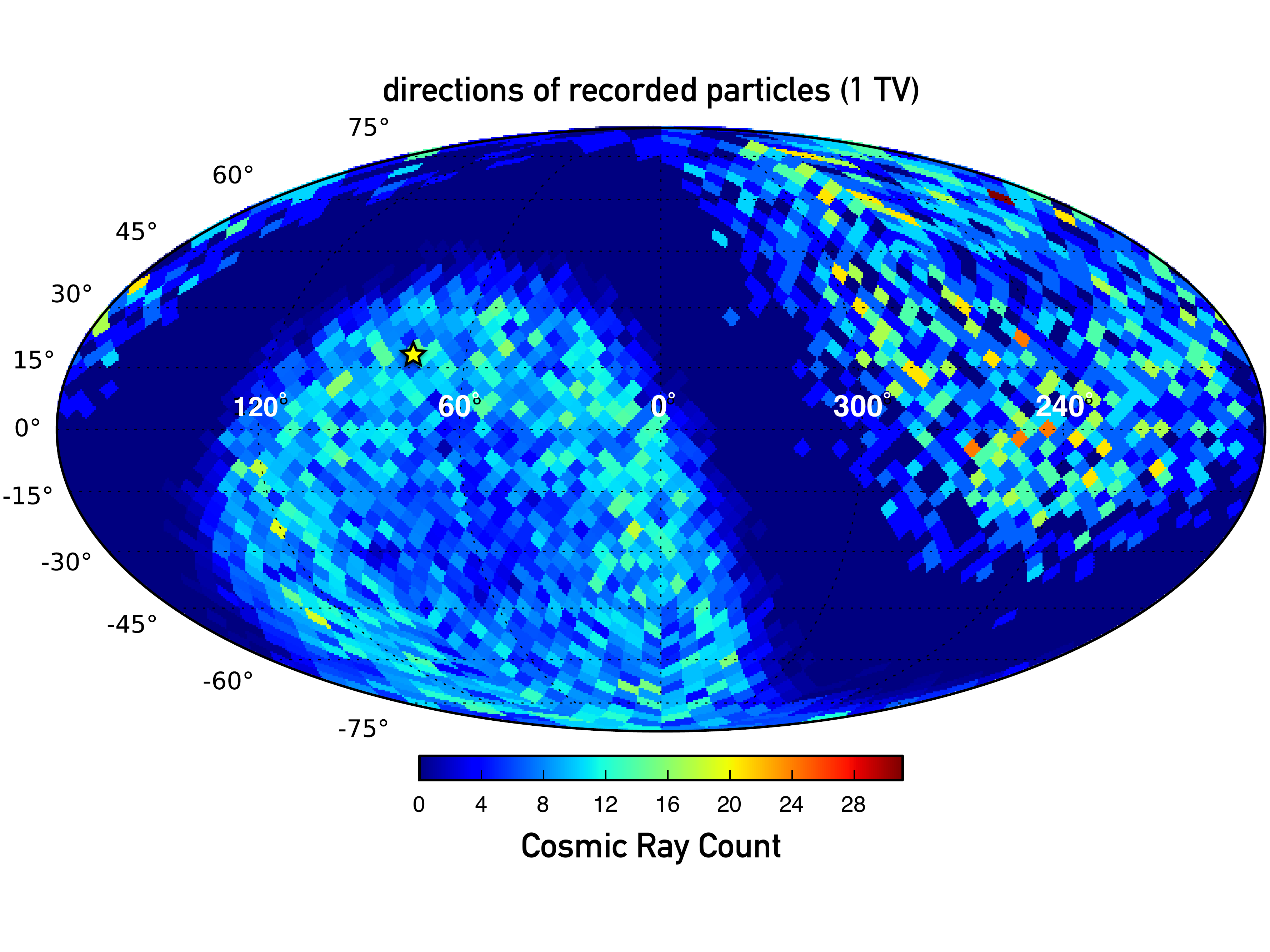}
\includegraphics[width=\columnwidth]{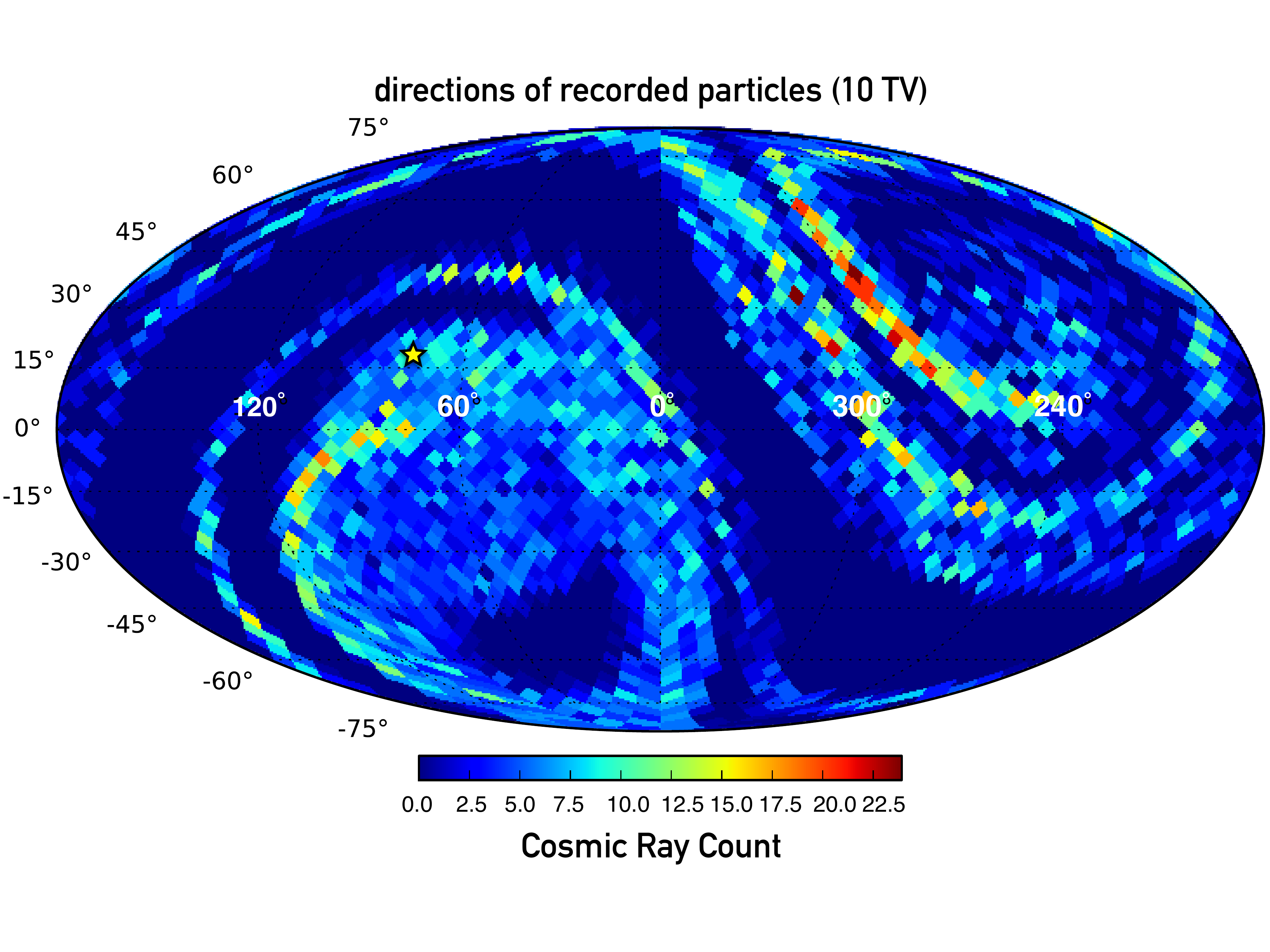}
\caption{{\it Top:} Map in equatorial coordinates of the positions of injected particles (from the 60$^{\circ}\times$60$^{\circ}$ region of the heliosphere upstream of the ISM flow). Only the initial positions of those particles that are actually recorded are shown here. {\it Center:} Map in equatorial coordinates of the arrival direction distribution of the recorded mixed composition particles at rigidity scale of 1 TV. {\it Bottom:} Map in equatorial coordinates of the arrival direction distribution of the recorded mixed composition particles at rigidity scale of 10 TV. The yellow star indicates the approximate position of the heliospheric tail. The dashed yellow box corresponds approximately with the region of initial position of all the particles.}
\label{fig:skymaps1}
\end{figure}
\begin{figure}[t!]
\includegraphics[width=\columnwidth]{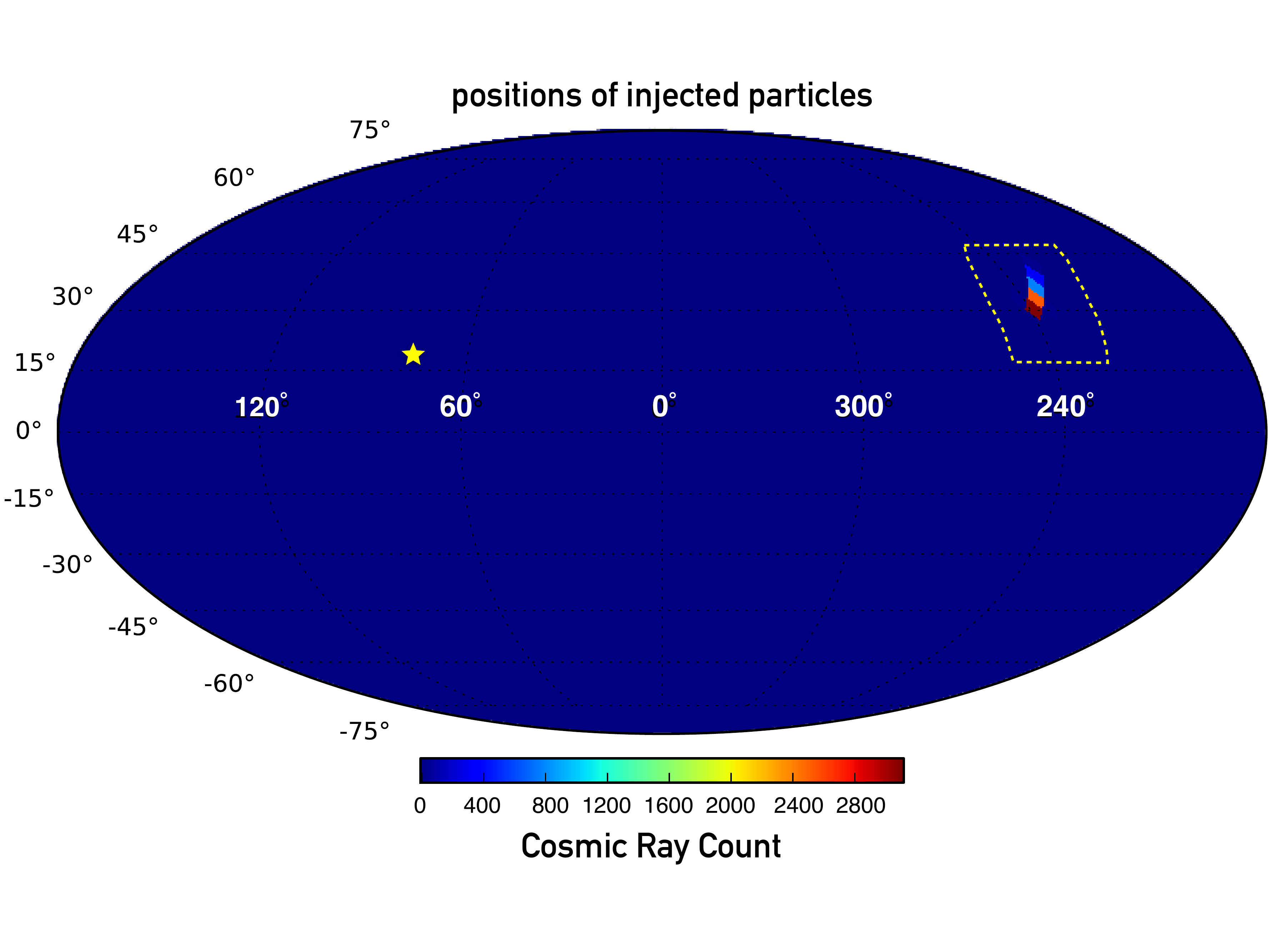}
\includegraphics[width=\columnwidth]{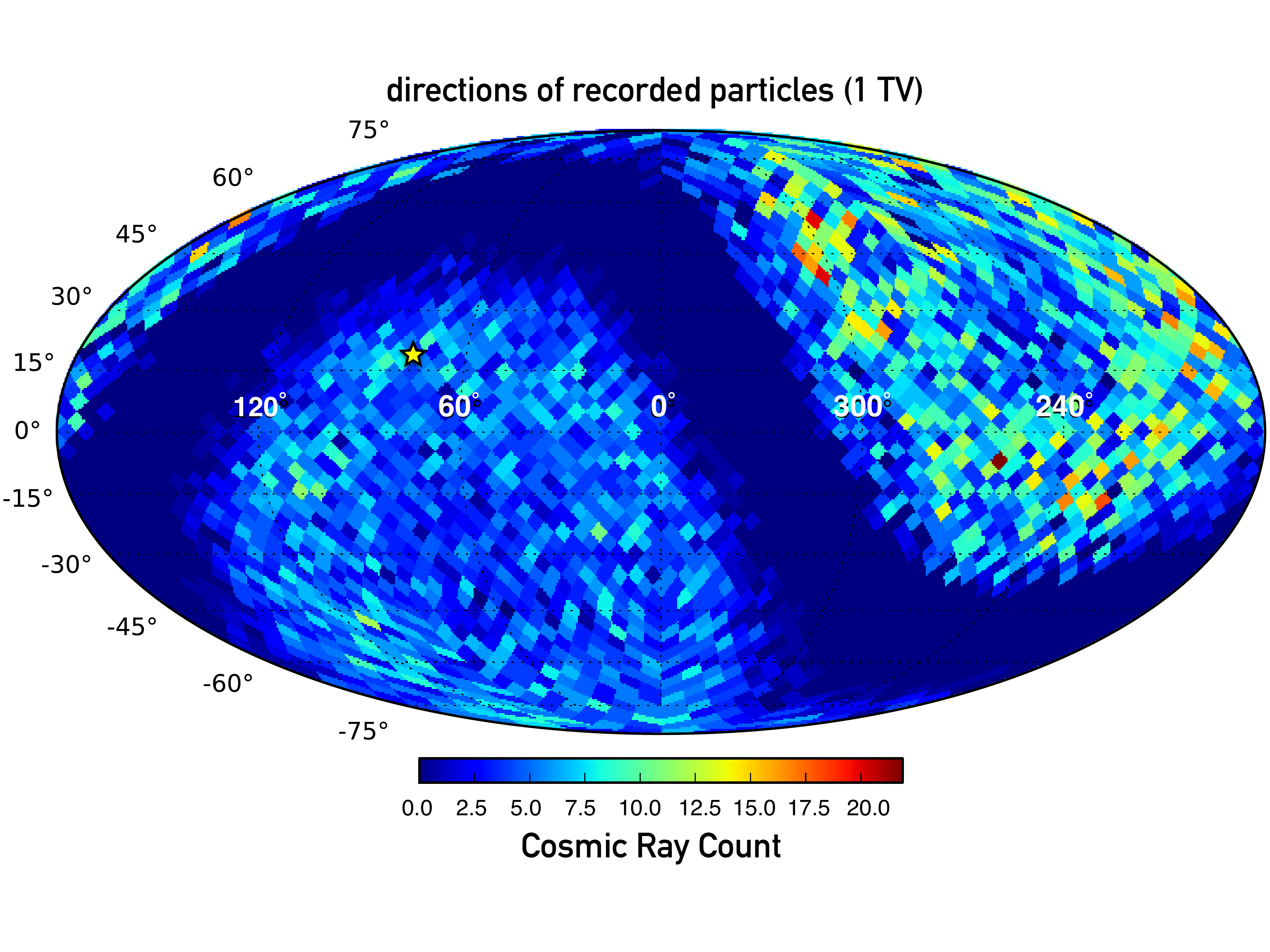}
\includegraphics[width=\columnwidth]{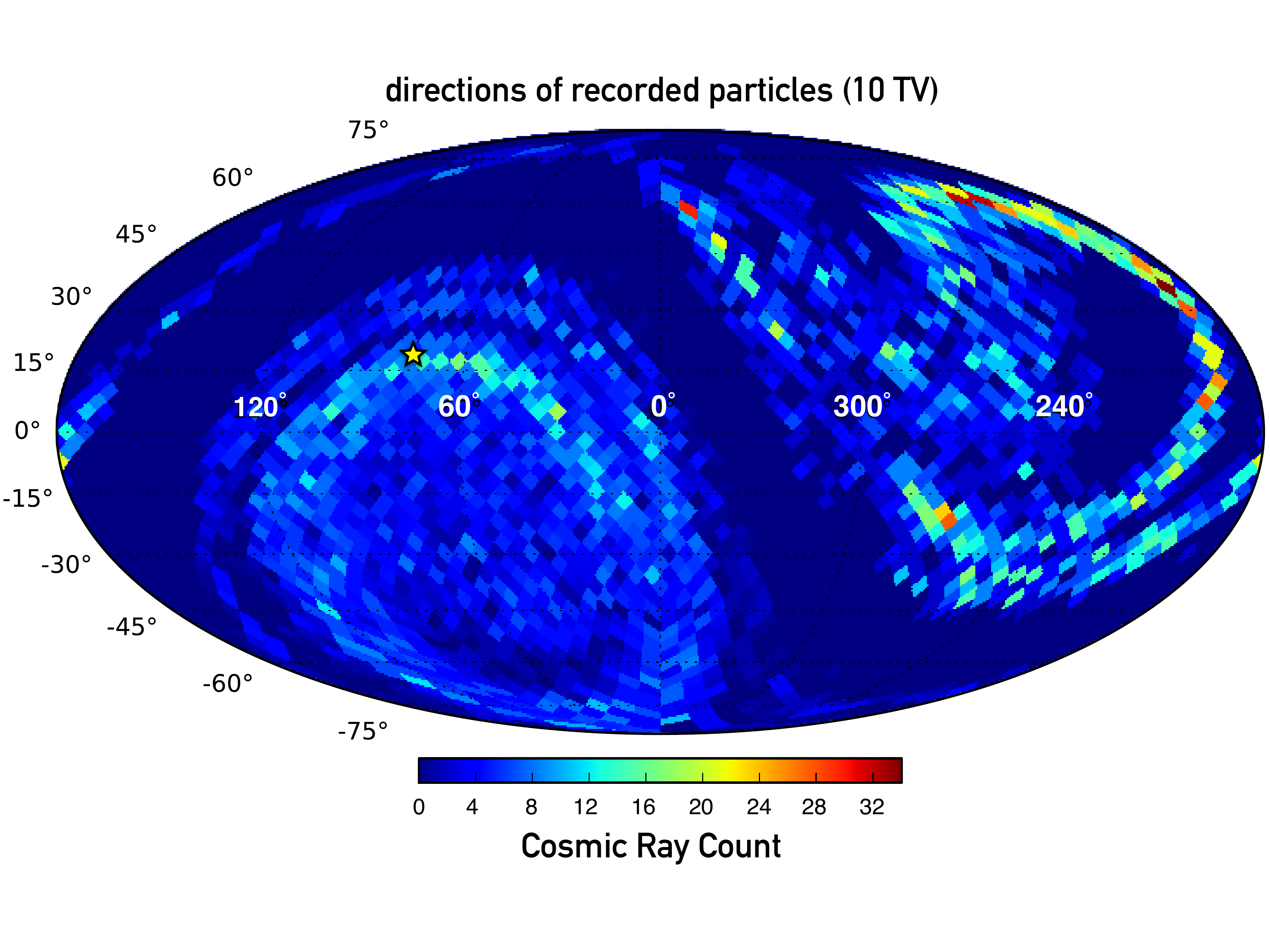}
\caption{{\it Top:} Map in equatorial coordinates of the positions of injected particles (from the 30$^{\circ}\times$30$^{\circ}$ zone of the heliosphere downstream of interstellar side, in proximity of the heliotail). Only the initial positions of those particles that are actually recorded are shown here. {\it Center:} Map in equatorial coordinates of the arrival direction distribution of the recorded mixed composition particles at rigidity scale of 1 TV. {\it Bottom:} Map in equatorial coordinates of the arrival direction distribution of the recorded mixed composition particles at rigidity scale of 10 TV. The yellow star indicates the approximate position of the heliospheric tail. The dashed yellow box corresponds approximately with the region of initial position of all the particles.}
\label{fig:skymaps2}
\end{figure}
The numerically integrated trajectories of the sets in Table~\ref{tab:sets} were combined according to the mixed CR composition from~\cite{gaisser_2013} (i.e., approximately 40\% protons, 40\% helium nuclei, and 20\% iron nuclei at low rigidity and 50\% protons and 50\% helium nuclei at high rigidity) and used to study the effects that scattering processes on the heliospheric magnetic field have on the particle arrival direction distributions. As mentioned in Section~\ref{sec:lt}, unlike the procedure followed in LX16, in this study particles were injected in two regions on the injection sphere at 6000 AU distance from Earth aligned along the LIMF and forward-propagated. At each point within those two regions, the particle directions were chosen from a uniform distribution towards the inner sphere. As shown in Table~\ref{tab:sets}, a large fraction of the injected particles does not reach the target sphere, in proximity of Earth, because of scattering processes in the magnetic instability regions, thus contributing to the anisotropic distribution independently of the initial CR density gradient.
\begin{figure*}[t!]
\begin{center}
\includegraphics[width=\columnwidth]{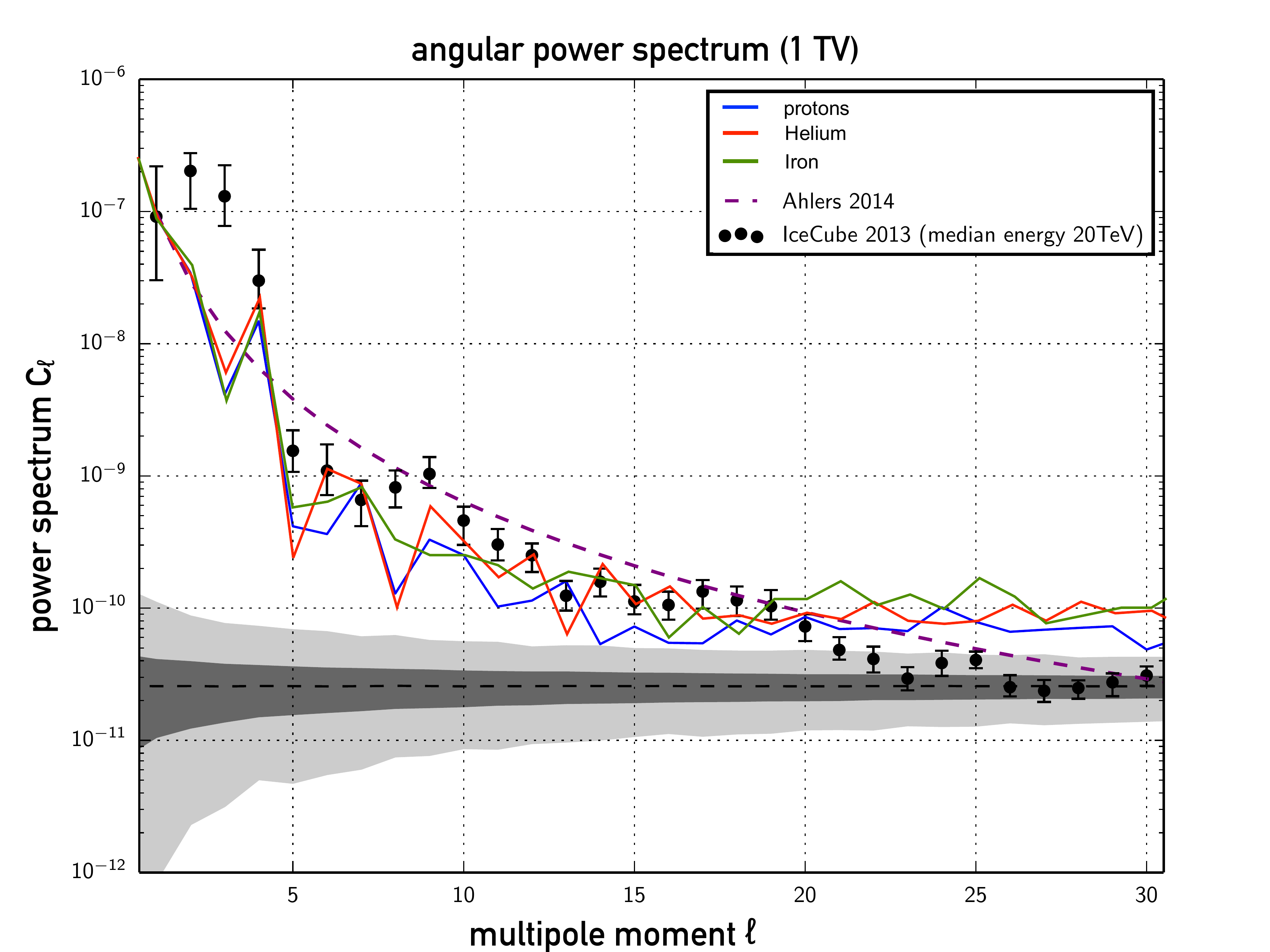}
\includegraphics[width=\columnwidth]{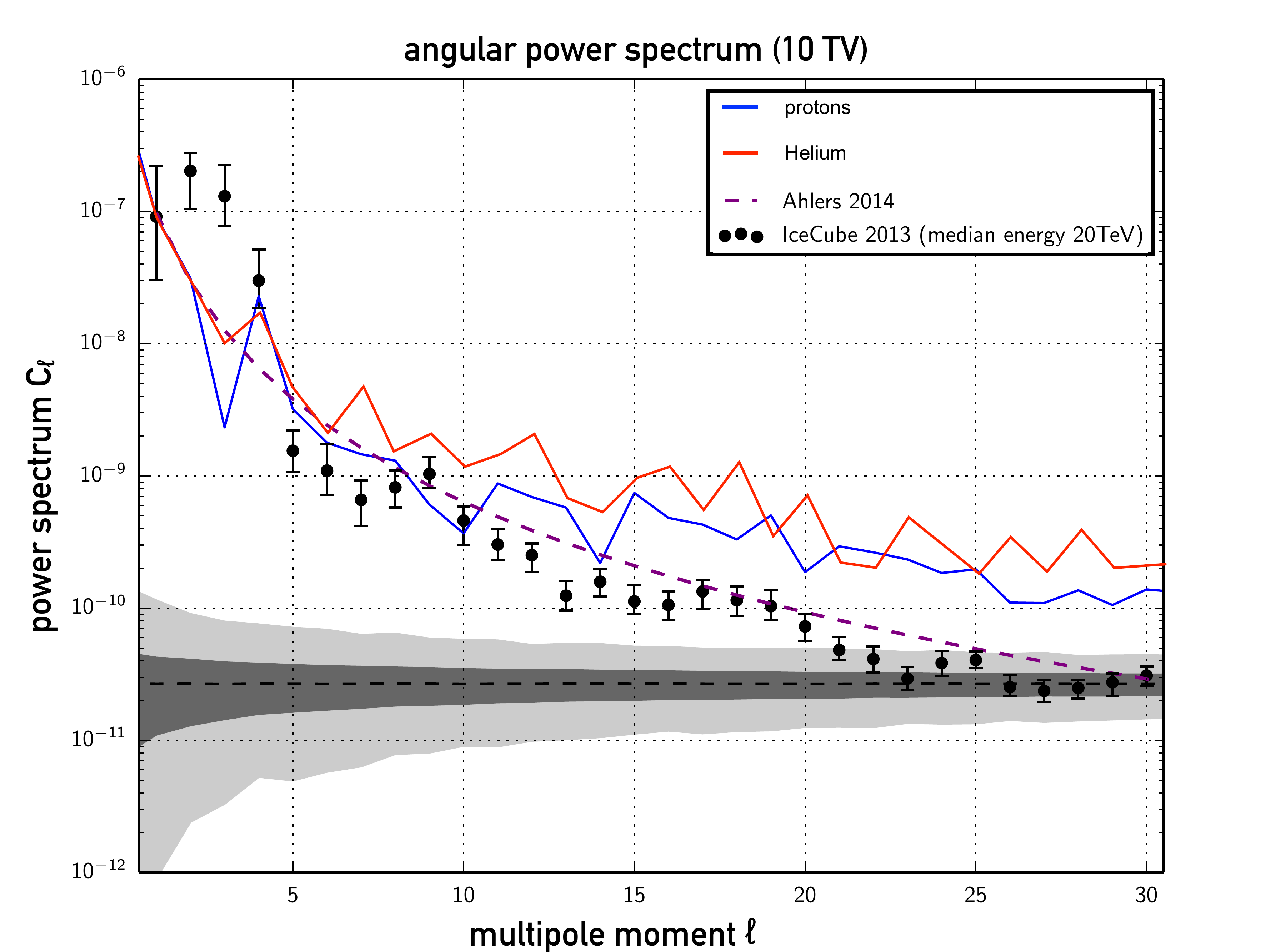}
\end{center}
\caption{Angular power spectrum of the arrival direction distribution on the target sphere of the 1-TV rigidity particle sets (on the left) and of the 10-TV rigidity particle sets (on the right). Protons (blue lines), helium nuclei (red lines), and iron nuclei (green line) are separately shown. The gray bands show the 1$\sigma$ and 2$\sigma$ bands for a large set of isotropic sky maps. The black circles are the results from the IceCube Observatory at a median energy of 20 TeV~\citep{santander_2013,aartsen_2015}. The dashed purple line is the power spectrum from~\cite{ahlers_2014}. The angular power spectrum results are normalized to the IceCube experimental results at the dipole component ($\ell$ = 1). Note that the angular power spectra are calculated with all particles initiated from both regions on the injection sphere.}
\label{fig:APS}
\end{figure*}

The particles hitting the target sphere were recorded and are represented in the sky maps of Figures~\ref{fig:skymaps1} (for particles injected in the region downstream of the ISM flow, and in proximity of the heliospheric tail) and~\ref{fig:skymaps2} (for particles injected in the region upstream of the ISM flow). In the figures, the direction of the heliotail is indicated with a yellow star.
At the top of the figures, the initial positions on the injection sphere of those trajectories that reach the target sphere are shown. The yellow dashed boxes indicate the region of initial positions of all generated particles. The limited size of those regions show that those particles streaming along the LIMF within a relatively narrow magnetic field-line tube have the highest chance of reaching the target sphere. Note that, within those localized regions, all particles have a wide range of uniformly distributed pitch angles (or directions), which determines the size of the corresponding gyroradius $r_L\approx (220 / Z)\,(E / TeV)\, \sqrt{1-\mu^2}\, (\mu G / B)$ AU, with $\mu$ the cosine of the pitch angle. The instantaneous gyroradius along the particle trajectories are shown Figure~\ref{fig:rl} and the corresponding mean and RMS values in Table~\ref{tab:sets}.
At the center of the figures, the arrival directions of the 1-TV rigidity scale particles at the target sphere are shown. The sky maps show that, although particles arrive at the heliosphere streaming along the LIMF from one specific direction, they are significantly redistributed. Approximately 50\% of the particles streaming from the downstream direction (and thus approaching the heliosphere in proximity of its elongated tail) undergo multiple scattering processes and appear as if they approach Earth from the upstream region (center of Figure~\ref{fig:skymaps1}). While, approximately 20-30\% of the particles streaming from the upstream direction (and thus approaching the heliosphere in the proximity of its nose) appear to approach Earth from the downstream region (center of Figure~\ref{fig:skymaps2}).
Resonant scattering along the heliotail has a more pronounced {\it focusing} effect towards the inner heliosphere. The wider injection region downstream (top of Figure~\ref{fig:skymaps1}) compared to that upstream (top of Figure~\ref{fig:skymaps2}) shows that the heliotail is able to {\it trap} particles more efficiently from larger distances downstream and collect them near Earth.

The arrival directions of the 10-TV rigidity scale particles on the target sphere are shown at the bottom of the figures. The effects of scattering on magnetic perturbations are still visible; however, the particle distribution is not as smooth, but it develops localized regions associated with the magnetic field geometric structure. Even at this high rigidity, it is possible to notice that multiple scattering effects of the heliotail are stronger for particles streaming downstream than for those upstream. At higher rigidity, with the decreasing influence of scattering, particle distribution is expected to converge to that at the injection sphere.

A side effect of the relatively large radius of the target sphere (200 AU, i.e., 2-3 times the gyroradius at 1-TV rigidity scale) is that, although the trajectories show the effect of multiple scattering processes across the heliospheric boundary (see Figure~\ref{fig:traj}), they are not propagated too deep into the vicinity of  Earth. As a consequence, although particles propagate across magnetic field lines with stochastically distributed pitch angle, the connection to the large-scale magnetic field direction is still relatively strong because of the general structure of the heliosphere. This generates very low particle populations along directions that are approximately perpendicular to the LIMF (visible as a dark band at the center and the bottom of Figures~\ref{fig:skymaps1} and~\ref{fig:skymaps2}).
%

The figures show a specific snapshot. However, particles reaching Earth are from all masses and energies, and ground-based experiments have relatively poor mass and energy resolutions. Each infinitesimal rigidity interval produces a characteristic fingerprint pattern similar to those in the figures, deeply dependent on the properties of the heliospheric magnetic field. 
Observations reveal the overlapping of those characteristic distributions, and, in fact, there is experimental evidence that coexisting anisotropy features originate from different energy ranges (see, e.g.,~\cite{bartoli_2013,aartsen_2015}). Even though in this study we provide a window on the possible effects of CR composition, a detailed prediction of the observation requires fine-tuning of several effects, which is not within the scope of this work.
\subsection{Angular Power Spectrum}
\label{sec:aps}

As discussed in the previous section, the topology of the sky maps in Figures~\ref{fig:skymaps1} and~\ref{fig:skymaps2} is associated with the specific rigidity scales used for the particular particle sets of Table~\ref{tab:sets}. Because of the stochastic nature of scattering processes, statistically uncorrelated data sets would produce similar but not identical sky maps. Even with a seemingly different topology, however, such sky maps would have the same angular power spectrum, since this contains  global statistical properties of the ensemble of particles, independent of the spatial location of the anisotropic features. The angular power spectrum, therefore, is a physically relevant quantity to study.
%

Figure~\ref{fig:APS} shows the angular power spectrum for the particle sets of Table~\ref{tab:sets}. On the left, the results for the 1-TV rigidity sets and on the right for the 10-TV rigidity sets. For each particle set, the trajectories propagating from both injection regions are used (see Section~\ref{ssec:sets}). As expected, the shape of the angular power spectrum depends on particle rigidity.  As in LX16, the figures shows the angular power spectrum from the IceCube observatory~\citep{santander_2013,aartsen_2015}, the power spectrum from~\cite{ahlers_2014}, and that corresponding to an isotropic arrival direction distribution of the same number of particles. In the figures, the power spectra are normalized to each other at the dipole component (i.e., $\ell$ = 1).


CR particles in the 1-TV rigidity scale, as discussed in Section~\ref{sec:skymaps}, are those most affected by multiple scattering induced by magnetic instabilities at the heliospheric boundaries. Their arrival direction distribution appears to develop angular structures on the order of 20$^{\circ}$ ($\ell \approx$ 10-15, where it reaches the statistical limit) as shown on the left of Figure~\ref{fig:APS}.
In the 10-TV rigidity scale, on the other hand, the small-scale filaments visible in Figures~\ref{fig:skymaps1} and~\ref{fig:skymaps2} contribute to the higher power for large values of $\ell$ on the right of Figure~\ref{fig:APS}.
As already mentioned, the sets used in this study represent two particular particle rigidity snapshots of the wide CR energy spectrum. Figure~\ref{fig:APS} highlights that a complex angular power spectrum arises even when particles propagate across relatively small-scale magnetic structures on a short distance scale. This overlaps with the effects of large-scale turbulence in the ISM over long distance scales (as discussed in LX16). While the ISM contribution is stochastically distributed, the heliospheric effects, although produced by scattering on magnetic instabilities, are expected to have directional correlations with the heliosphere. The experimental separation of the two contributions is the key to exploring the properties of the heliospheric magnetic field with TeV CRs.
%

\section{Discussion}
\label{sec:disc}


We have shown the dramatic effects that the heliosphere imprints on the CR arrival distribution at Earth. Our results show that the interactions of CRs with the heliosphere are relevant, and for the 1-10 TV range, cosmic-ray arrival cannot be studied without taking these effects into account. 


The feature that distinguishes this work from previous studies is the forward-propagation technique used. We have shown that Liouville's theorem cannot be applied in this case because resonant scattering on magnetic perturbations on the heliospheric magnetic field generates stochastic pitch angle distributions, and it effectively breaks adiabaticity (Section~\ref{sec:lt}); therefore a back-propagation approach is not applicable. Another important idea behind the forward propagation is the concept of particle escape. As pointed out in Section~\ref{ssec:crprop} and in~\cite{rettig_pohl_2015}, particle escape due to transport in complex magnetic fields contributes to the resulting arrival direction distribution. 
Those particles that escape without ever reaching a given target location build up an uneven arrival direction distribution. In a back-propagation approach, there is no particle escape by construction. In this case, a weight function is used to inject an initial anisotropy for the reversed trajectories. The final anisotropy results from the spatial redistribution of a constant particle density. 

The results of the present study are different from those in~\cite{schwadron_2014}, where the importance of a relative large-scale excess of CR (due to a nearby supernova) was highlighted, without accounting for the effects of scattering on the heliospheric magnetic field. In our study, we use a self-consistent solution to the ideal MHD equations to obtain a model of the heliosphere. In that way, we can assess the direct heliospheric effects on the CR arrival distribution.
And our approach is fundamentally different from the study in~\cite{zhang_2014} and~\cite{schwadron_2014}, where a back-propagation approach was used. In the case of~\cite{zhang_2014}, a model of the heliosphere was used, although different from the one used in the present study in that magnetic instabilities were not as prominent.  Even though our forward propagation approach is computationally expensive, it is the only acceptable technique since the conditions of Liouville's theorem are not satisfied given the effective collision interactions with the instabilities in the heliosphere (Section~\ref{sec:lt}). 

%
Transport across the galactic magnetic field may be described with homogeneous anisotropic diffusion scenarios, where particles propagate faster along the magnetic field lines. If diffusion describes large-scale CR propagation in the ISM, the arrival direction distribution is expected to have a dipolar shape oriented along the LIMF. Even in the presence of multiple angular scale structures from nondiffusive propagation effects (see Section~\ref{sec:intro}), it is usually assumed that the dipole component still has a direct connection to the underlying large-scale diffusion. Observations show that the dipole component appears to be aligned with the LIMF~\citep{ahlers_2016} after accounting for the experimental biasing projection of angular features on the equatorial plane and the limited field of view of all ground-based experiments~\citep{ahlers_etal_2016}.
The actual direction of the dipole component can be considered accidental and perhaps associated to a relatively recent and nearby source of CRs contributing to the excess on one side of the LIMF lines rather than the other (see, e.g.,~\cite{schwadron_2014} and~\cite{ahlers_2016}).
%

In reality, the structured interstellar magnetic field, with its different overlapping contributions at different spatial scales, easily generates deviations from the simple diffusion scenario. Particles with a given rigidity are likely to be more strongly affected by magnetic fields with gyro-scale spatial structures. This can cause dramatic changes in the transport properties (see for instance ~\cite{desiati_zweibel_2014} and references therein). At a given rigidity scale, CR particle distributions are shaped, over their entire propagation history, by the accumulating effects of magnetic perturbations at scales $\ll r_L$, by the magnetic field geometric structure at scales $\gg r_L$, and by the strong resonant effects on geometric features at scale $\approx r_L$. The LIMF is found to be coherent within approximately 60 parsec~\citep{frisch_2012}. This means that CRs approach the heliosphere from the ISM spiraling around the LIMF lines with pitch angle distribution reflecting their propagation history.
At 1-TV rigidity scale, it is, therefore, expected that the heliosphere, with its approximately gyro-scale magnetic instabilities, has the power to redistribute CRs. 
The strong heliospheric influence on the arrival direction distribution of CR particles is shown in Figures~\ref{fig:skymaps1} and~\ref{fig:skymaps2}, where a large fraction of the particles passing through the heliosphere is redistributed in pitch angle by multiple scattering in the magnetic instabilities. While the global anisotropy may still be ordered by the LIMF, the medium- and small-scale features depend on the peculiar properties of the heliospheric magnetic field. In the 1-TV rigidity scale, where resonant scattering processes are more dominant, particles are severely redistributed, and this results in a relatively smoother distribution. On the contrary, in the 10-TV rigidity scale, scattering processes are still significant, but the large scale average magnetic field induces the formation of medium- and small-scale structures.

In other words, the heliosphere acts as a {\it diffusor} where particles with gyroradius $r_L \approx L_{\text{instabilities}} \approx 10-100$ AU are more stochastically influenced via resonant multiple scattering than those with $r_L \gtrsim L_{\text{heliosphere}} \approx 600$ AU. 
%
A consequence of violation of magnetic moment conservation (see Figure~\ref{fig:promap}) is that an initially uniform pitch angle distribution becomes more structured with most of the energy stored in the large angular scales. The lower the effect of stochastic redistribution from scattering, the higher the contribution from small angular scale structures, as shown in Figure~\ref{fig:APS}. It is likely that the largest angular scale, such as dipole and quadrupole, are affected by these scattering processes as well. However, in this work, such an effect is not explicitly assessed, since particles need to be propagated much closer to Earth. 

Statistically uncorrelated data sets can produce sky maps that have similar features; nonetheless, they will not be identical since these scattering processes are stochastic by nature. These sky maps may look different but they share the same angular power spectrum, since it encloses the global statistical properties of the ensemble, and not the specific spatial locations of the maps' features. Consequently, the physically relevant quantity is the angular power spectrum.

The energy transfer between angular scales is different from that studied in LX16, where the effect of turbulence on a back-propagated particle distribution was considered. In that case, the particle gyroradius $r_L$ was always smaller than the largest scale of magnetic turbulence. Therefore, as long as $r_L < L_{\text{injection}}$, particles are always in resonance with a given turbulence scale, the largest scale conveying more power than smaller scales. The energy of an initially dipolar pitch angle distribution is more rapidly transferred into smaller angular scales at higher energy, because resonant scattering occurs on higher power turbulence scales. In both studies, LX16 and this work, it is found that magnetic scattering generates flatter power angular spectra at higher rigidity scales. In general, if scattering is sufficiently strong, it is possible to form small-scale features within a relatively short distance scale as well.
%
%
In the heliospheric model used in the present study, the size and location of the magnetic instabilities mostly influence 1 TV-scale particles, especially those propagating from downstream of the interstellar wind, i.e., in closer proximity to the elongated tail.
It is important to note that other heliospheric properties, not accounted for in the model used here, may significantly increase the scattering rate at a specific rigidity range. The 11-year solar magnetic field inversion cycle generates pockets of magnetic polarity dragged outward and along the heliotail by solar wind~\citep{pogorelov_2006}. The size of those regions is approximately 200-300 AU, as discussed in DL13, which resonate with particles with a rigidity of a few TV. Since the geometry of such polarity regions is different than that of magnetic instabilities, the effect can account for different distributions. This effect was not considered in the present study and will be the subject of a followup project.

One carefully studied possibility is that the nonconservation of the magnetic moment discussed in Section~\ref{sec:lt} may result from poor accuracy of the trajectory interaction code used.
%
%
%
The numerical algorithm used to integrate particle trajectories in this study is the same used in LX16, in the study of MHD turbulence effect of the particle distribution. As extensively discussed in LX16, the Bulirsch-Stoer numerical integration method used is considered one of the best known algorithms satisfying both high accuracy and efficiency~\citep{press_1986} and widely applied in the literature~\citep{giacalone_jokipii_1999,xu_yan_2013}. The accuracy of the numerical integration is further controlled by monitoring the local truncation error estimated at each time step. If the relative error is larger than the relative tolerance level of 10$^{-6}$, the step size is adaptively reduced in order to limit the error accumulation in both momentum and spatial coordinates, across the maximum integration time used in this study (between 10,000 and 100,000 gyrations). The accuracy in momentum coordinates was tested, and the resulting energy conservation was found to be constrained well within 10$^{-5}$ (see LX16). Due to the adaptive step size, the accuracy in spatial coordinates is $\ll$ r$_L$, thus limiting numerical diffusion to an undetectable level. In LX16, the violation of magnetic moment was not statistically relevant (and therefore the back-propagation method was utilized), while in the current heliospheric study it is dramatically significant. The difference between the two cases is in the magnetic field used. The numerical accuracy of the trajectory integration method is the same. The magnetic field interpolation via 3D cubic spline functions does not appear to reduce numerical accuracy.

As described in LX16, it is possible to exclude with confidence that the particle trajectory integration method used in this study can induce chaotic behavior as a result of poor accuracy. Therefore it is the opinion of the authors that the effective collisional behavior found in this case is due to the properties of the heliospheric magnetic field used. 

To conclude, the study presented in this work explores the effects of the heliosphere on multi-TeV CR arrival direction distribution. The results illustrate the importance of particle rigidity, $E/Ze$, in making it possible that the heliospheric influence stretches across a relatively wide CR particle energy range. For the particular model of the heliosphere by~\cite{pogorelov_2015} (described in Section~\ref{sec:heliomf}), resonant scattering processes are strong enough to break adiabaticity and generate stochastic pitch angle distributions. As a consequence, Liouville's theorem could not be applied and a computationally expensive forward propagation technique was used. The escape of particles due to magnetic bottle mirroring and multiple resonant scattering generates a rigidity-dependent complex arrival distribution of CR particles that comprises a wide power spectrum in angular structures. 
In order to reproduce the observations, proper consideration must be given to a wider range in rigidity accounting for the actual spectrum and composition of the CRs, especially taking into account that the precise features are not exactly reproducible given the stochastic process at the heliospheric boundary, and an angular power spectrum approach should be taken in order to study the CR-arrival anisotropy. 
%
%
Assuming that the dipole component of the observed CR anisotropy is the imprint of diffusion in the ISM, the study of the complex angular structure can provide important hints as to the turbulent properties of the ISM (especially at energies $>$ 100 TeV, as shown in LX16) and to the properties of the heliosphere (in the TeV energy range).


\section{Summary}
\label{sec:concl}

%

The main results can be summarized as follows:
\begin{itemize}
\item As CRs are strongly affected by magnetic structures on the order of their gyroradius, multi-TeV particles are subject to significant heliospheric scattering. This redistributes CRs and affects their arrival direction distribution. Our work shows that this scattering can have a significant effect. 

\item Our simulations show significant resonant scattering of the CRs by the heliosphere. Therefore, the conditions of Liouville's theorem are not satisfied and the backward-propagation technique cannot be used to study CR 
anisotropies arising from the interaction of heliospheric magnetic fields with CRs.

\item Our study of the effect of the heliosphere on CR anisotropy, which we performed using the forward-propagation techniques, demonstrates the following features:
\begin{itemize}
\item Results for protons: The heliosphere has a strong effect in redistributing CRs at the 1 TV rigidity scale. Multiple scattering with stochastic pitch angle redistribution is relevant, and anisotropy arises even without assuming any weight in the initial arrival distribution of CR particles. The scattering effect is weaker at the 10 TV rigidity scale but is still important in producing small- to medium-angular scale features that contribute to the overall arrival direction distribution.
\item  Results for heavy nuclei:  At 10 TeV, the flux of helium nuclei starts to become dominant, while heavy nuclei contribute to about a third of the total flux~\citep{gaisser_2013}. We found that high-energy heavy nuclei have an important role in shaping the observed anisotropy at very small angular scales (i.e., multipole moments $\ell$ = 7-25). This also means that the heliospheric influence affects observed anisotropies over a wide energy range.
\end{itemize}

\item Our study calls for both more extensive observations of CR anisotropies and more detailed numerical testing using high-resolution models of the heliosphere. Future research should also take into account the significant time variations of the heliospheric magnetic field related to both the 11-year cycle and dynamical instabilities on the longer time scales. 

\end{itemize}

\acknowledgments

The authors wish to thank colleagues at WIPAC and the Department of Astronomy for discussions on CR anisotropy. P.D. acknowledges support from the U.S. National Science Foundation Office of Polar Programs. A.L. acknowledges support of NSF grant AST-1212096 and NASA grant X5166204101. A.L. and V.L.B. acknowledge support from NASA grant NNX14AJ53G.
N.P. was supported by NASA grants NNX14AJ53G, NNX14AF43G, and NNX15AN72G, DOE grant DE-SC0008334, and NSF Petascale Computing Resource Allocation award ACI-1615206.

%
%


\begin{thebibliography}{}


\bibitem[Aartsen et al.(2013)]{aartsen_2013}
Aartsen, M. et al. 2013b, Astrophys. J. 765, 55

\bibitem[Aartsen et al.(2016)]{aartsen_2015}
Aartsen, M. et al. 2016, accepted for publication on ApJ

\bibitem[Abbasi et al.(2010)]{abbasi_2010}
Abbasi, R. et al. 2010a, Astrophys. J. 718, L194

\bibitem[Abbasi et al.(2011)]{abbasi_2011}
Abbasi, R. et al. 2011a, Astrophys. J. 740 16

\bibitem[Abbasi et al.(2012)]{abbasi_2012}
Abbasi et al. 2012b, Astrophys. J. 746, 33

\bibitem[Abdo et al.(2008)]{abdo_2008}
Abdo, A.A. et~al. 2008, Phys. Rev. Lett. 101, 221 101

\bibitem[Abdo et al.(2009)]{abdo_2009}
Abdo, A.A. et~al. 2009, Astrophys. J. 698, 2121

\bibitem[Abeysekara et al.(2014)]{abeysekara_2014}
Abeysekara, A.U. et al. 2014, Astrophys. J. 796, 108

\bibitem[Aglietta et al.(2009)]{aglietta_2009}
Aglietta, M. et~al. 2009, Astrophys. J. 692, L130

\bibitem[Ahlers(2014)]{ahlers_2014}
Ahlers, M. 2014 Phys. Rev. Lett. 112, 021101

\bibitem[Ahlers et al.(2016a)]{ahlers_etal_2016}
Ahlers, M. et al. 2016, ApJ 823, 10

\bibitem[Ahlers(2016b)]{ahlers_2016}
Ahlers, M. 2016 arXiv:1605.06446

\bibitem[Ahlers \& Mertsch(2015)]{ahlers_mertsch_2015}
Ahlers, M., \& Mertsch, P. 2015, arXiv:1506.05488

\bibitem[Amenomori et al.(2005)]{amenomori_2005}
Amenomori, M. et~al. 2005, Astrophys. J. Lett. 626, L29

\bibitem[Amenomori et al.(2006)]{amenomori_2006}
Amenomori, M. et~al. 2006, Science, 314, 439

\bibitem[Amenomori et al.(2011)]{amenomori_2011}
Amenomori, M. et~al. 2011, Proc. 32nd ICRC, Beijing China

\bibitem[Amenomori et al.(2007)]{amenomori_2007}
Amenomori, M. et~al. 2007, Proc. 30th ICRC, M\'erida, Mexico

\bibitem[Avinash et al.(2014)]{Avinash}
Avinash, K., Zank, G. P., Dasgupta, B., Bhadoria, S.~2014,
\apj, 791, 102

\bibitem[Bartoli et al.(2013)]{bartoli_2013}
Bartoli, B., et~al. 2013, Phys. Rev. D 88-8, 082001

\bibitem[Bartoli et al.(2015)]{bartoli_2015}
Bartoli, B., et~al. 2015, Astrophys. J. 809, 90

\bibitem[Baumjohann \& Treumann(1996)]{baumjohann_1996}
Baumjohann, W. \& Treumann, R.~A. 1996, Plasma Physics, Magnetohydrodynamics, Kinematics, Radiative Transfer, London: Imperial College Press

\bibitem[Belov \& Ruderman(2010)]{Belov}
Belov, N.~A, \& Ruderman, M.~S.~2010,
\mnras, 401, 607

\bibitem[Beresnyak et al.(2011)]{beresnyak_2011}
Beresnyak, A., Yan, H., \& Lazarian, A. 2011, Astrophys. J. 728, 60, 8 pp.

\bibitem[Biermann et al.(2013)]{biermann_2013}
Biermann, P.L., Becker Tjus, J., Seo, E.-S., \& Mandelartz, M. 2013, Astrophys. J. 768, 124

\bibitem[Blasi \& Amato(2012)]{blasi_2012}
Blasi, P., \& Amato, E. 2012, JCAP 1, 11

\bibitem[Borovikov et al.(2008)]{Borov08}
Borovikov, S.~N., et al. 2008, Astrophys. J. 682, 1404

\bibitem[Borovikov \& Pogorelov(2014)]{Borov14}
Borovikov, S.~N., Pogorelov, N.~V.~2014,
\apj, 783, LX16


\bibitem[Bradt(2008a)]{bradt_2008a}
Bradt, H. 2008, Astrophysics processes, Cambridge Univ. Press, Cambridge

\bibitem[Bradt(2008b)]{bradt_2008b}
Bradt, H. 2008, in Astrophysics Process (Cambridge, (supplement): Cambridge Univ. Press)

\bibitem[Brandenburg \& Lazarian(2013)]{branden_lazarian_13}
Brandenburg, A. \& Lazarian, A. Space Sci Rev (2013) 178: 163.

\bibitem[Burkhart et al.(2014)]{burkhart_2014}
Burkhart, B. et al. 2014, Astrophys. J. 790 130

\bibitem[Cesarsky(1980)]{cesarsky_1980}
Cesarsky, C. J. 1980, ARA\&A, 18, 289

\bibitem[Chalov(1996)]{Chalov}
Chalov, S.~V.~1996,
\aap, 308, 995

\bibitem[Chandran(2000)]{chandran_2000}
Chandran, B. D. G. 2000, PhRvL, 85, 4656

\bibitem[Cho \& Lazarian(2002)]{cho_lazarian_2002}
Cho, J., \& Lazarian, A. 2002, Phys. Rev. Lett. 88, 245001

\bibitem[Cho \& Lazarian(2003)]{cho_lazarian_2003}
Cho, J., \& Lazarian, A. 2003, MNRAS, 345, 325

\bibitem[de Jong et al.(2011)]{dejong_2011}
de Jong, J. et al. 2011, Proc. 32nd ICRC, Beijing, China

\bibitem[Desiati \& Lazarian(2012)]{desiati_lazarian_2012}
Desiati, P. \& Lazarian, A. 2012, NPG 19, 351

\bibitem[Desiati \& Lazarian(2013)]{desiati_lazarian_2013}
Desiati, P. \& Lazarian, A. 2013, Astrophys. J. 762, 44

\bibitem[Desiati \& Zweibel(2014)]{desiati_zweibel_2014}
Desiati, P. \& Zweibel, E.G. 2014, Astrophys. J. 791, 51

\bibitem[Drake et al.(2006)]{drake_2006}
Drake, J. F., Swisdak, M., Che, H., \& Shay, M. A. 2006, Nature, 443, 553

\bibitem[Effenberger et al.(2012)]{effenberger_2012}
Effenberger, F. et al. 2012, A\&A 547, A120

\bibitem[Erlykin \& Wolfendale(2006)]{erlykin_2006}
Erlykin A.D., \& Wolfendale A.W. 2006, Astropart. Phys. 25, 183

\bibitem[Farmer \& Goldreich(2004)]{farmer_gold_2004}
Farmer, A. J., \& Goldreich, P. 2004, ApJ, 604, 671

\bibitem[Florinski et al.(2005)]{Florin05}
{Florinski}, V., {Zank}, G.~P., \& {Pogorelov}, N.~V. 2005, J. of Geophys. Res. (Space Physics), 110, 7104

\bibitem[Florinski et al.(2013)]{florinski_2013}
Florinski, V, Jokipii, J. R., Alouani-Bibi, F., le Roux, J. A. 2013, Astrophys. J. 776, L37

\bibitem[Florinski et al.(2015)]{florinski_2015}
Florinski, V., Stone, E.C., Cummings, A.C., \& le Roux, J.A. 2015,  Astrophys. J. 803 47

\bibitem[Frisch et al.(2012)]{frisch_2012}
Frisch, P.C. et al. 2012 Astrophys. J. 760, 106

\bibitem[Frisch et al.(2014)]{frisch_2014}
Frisch, P.C. 2015 J. Phys.: Conf. Ser. 577, 012010

\bibitem[Gaensler et al.(2011)]{gaensler_2011}
B. M. Gaensler, et al. 2011, Nature 478, 214

\bibitem[Gaisser et al.(2013)]{gaisser_2013}
Gaisser, T.K., Stanev, T. \& Tilav, S. 2013 Front. of Phys. 8-6, 748

\bibitem[Galtier et al.(2000)]{galtier_2000} 
Galtier, S., Nazarenko, S.~V., Newell, A.~C., \& Pouquet, A. 2000, Journal of Plasma Physics, 63, 447

\bibitem[Giacinti \& Sigl(2012)]{giacinti_sigl_2012}
Giacinti, G., \& Sigl, G., 2012 Phys. Rev. Lett. 109, 071101

\bibitem[Giacalone \& Jokipii(1999)]{giacalone_jokipii_1999}
Giacalone, J. \& Jokipii, J.R. 1999, Astrophys. J. 520, 204

\bibitem[Goldreich \& Sridhar(1995)]{gs_1995}
Goldreich, P., \& Sridhar, S. 1995, Astrophys. J. 438, 763 

\bibitem[Goldstein et al.(2002)]{goldstein_2002}
Goldstein, H., Poole, C., \& Safko, J. 2002, Classical Mechanics (3rd ed.; San Francisco, CA: Addison-Wesley)

\bibitem[Gorski et al.(2005)]{gorski_2005}
Gorski, K. M., Hivon, E., Banday, A. J., et al. 2005, ApJ, 622, 759

\bibitem[de Gouveia dal Pino \& Lazarian(2005)]{deg_lazarian_2005}
de Gouveia dal Pino, E.~M. \& Lazarian, A. 2005, A\&A, 441, 845.

\bibitem[Guillian et al.(2007)]{guillian_2007}
Guillian, G. et al. 2007, Phys. Rev. D 75, 062003

\bibitem[Hall et al.(1999)]{hall_1999}
Hall, D.L. et~al. 1999, J. of Geophys. Res. 104, 6737

\bibitem[Haverkorn et al.(2008)]{haverkorn_2008}
Haverkorn, B., et al. 2008, Astrophys. J. 680, 362

\bibitem[Izmodenov \& Alexashov(2003)]{Izmod03}
Izmodenov, V. V., Alexashov, D. B.~2003,
Astronomy Lett., 29, 58

\bibitem[Kozai et al.(2014)]{kozai_2014}
Kozai et al. 2014, Earth, Planets and Space  66, 151

\bibitem[Kowal et al.(2009)]{kowal_2009}
Kowal, G. et al. 2009, Astrophys. J. 700, 63

\bibitem[Kowal et al.(2012)]{kowal_2012}
Kowal, G. et al. 2012, Nonlinear Processes in Geophysics, 19, 297

\bibitem[Kumar \& Eichler(2014)]{kumar_2014}
Kumar, R., \& Eichler, D. 2014 Astrophys. J. 785, 129

\bibitem[Jansson \& Farrar(2012a)]{jansson_farrar_2012a}
Jansson, R., \& Farrar, G.R. 2012a, Astrophys. J. 757, 14

\bibitem[Jansson \& Farrar(2012b)]{jansson_farrar_2012b}
Jansson, R., \& Farrar, G.R. 2012b, Astrophys. J. 761, L11

\bibitem[Jokipii(1966)]{jokipii_1966}
Jokipii, J.R. 1966, Astrophys. J. 146, 480

\bibitem[Jokipii \& Parker(1969)]{jokipii_parker_1969}
Jokipii, J.R. \& Parker, E.N. 1969, Astrophys. J. 155, 777

\bibitem[Lazarian(2006)]{lazarian_2006}
Lazarian A., 2006 AIPC, 874, 301

\bibitem[Lazarian(2007)]{lazarian_2007}
Lazarian, A. 2007, J. Quant. Spectros. \& Radia. Transfer 106, 225

\bibitem[Lazarian(2016)]{lazarian_2016}
Lazarian, A. 2016, arXiv:1607.02042

\bibitem[Lazarian \& Beresnyak(2006)]{lazarian_beresnyak_2006}
Lazarian, A., \& Beresnyak, A. 2006, MNRAS, 373, 1195

\bibitem[Lazarian \& Desiati(2010)]{lazarian_desiati_2010}
Lazarian, A. \& Desiati, P. 2010, Astrophys. J. 722, 188

\bibitem[Lazarian \& Vishniac(1999)]{lv_1999}
Lazarian, A., \& Vishniac, E.T. 1999, Astrophys. J. 517, 700

\bibitem[Lazarian \& Yan(2014)]{lazarian_yan_2014}
Lazarian, A., \& Yan, H. 2014, ApJ, 784, 38

\bibitem[Lazarian et al.(2015)]{lazarian_2015}
Lazarian, A., Eyink, G., Vishniac, E., \& Kowal, G. 2015, RSPTA, 373, 40144

\bibitem[Lithwick \& Goldreich(2001)]{lg_2001}
Lithwick, Y., \& Goldreich, P. 2001 Astrophys. J. 562, 279

\bibitem[Liewer et al.(1996)]{liewer96}
Liewer, P.~C., {Karmesin}, S.~R., \& {Brackbill}, J.~U. 1996, J. of Geophys. Res., 101, 17119


\bibitem[Longair(2011)]{longair_2011}
Longair, M. 2011, High Energy Astrophysics (3rd ed.; Cambridge: Cambridge Univ. Press)

\bibitem[L{\'o}pez-Barquero et al.(2016)]{barquero_2015}
{L{\'o}pez-Barquero}, V. et al.~2016,
\apj, 830, 19


\bibitem[Mertsch \& Funk(2015)]{mertsch_2015}
Mertsch, P. \& Funk, S. 2015 Phys. Rev. Lett. 114, 021101

\bibitem[Minnie et al.(2009)]{minnie_2009}
Minnie, J., et al. 2009, JGR 114, A01102

\bibitem[Munakata et al.(2010)]{munakata_2010}
Munakata, K. et~al. 2010, Astrophys. J. 712, 1100

\bibitem[Nagashima et al.(1988)]{nagashima_1998}
Nagashima, et~al. 1998, J. of Geophys. Res. 1031, 17429

\bibitem[Manuel et al.(2014)]{manuel_2014}
Manuel, R., Ferreira, S., \& Potgieter, M. 2014, Solar Physics 289, 2207

\bibitem[Opher et al.(2015)]{Opher15}
Opher, M., Drake, J. F., Zieger, B., Gombosi, T. I.~2015,
\apj, 800, 28

\bibitem[Pogorelov et al.(2006)]{pogorelov_2006}
Pogorelov, N.V., Zank, G.P., \& Ogino, T. 2006, Astrophys. J. 644, 1299

\bibitem[Pogorelov et al.(2013)]{pogorelov_2013}
Pogorelov, N.V., Suess, S.T., \& Borovikov, S.N. 2013, Astrophys. J. 772, 2

\bibitem[Pogorelov et al.(2015)]{pogorelov_2015}
Pogorelov, N.V., et~al. 2015, Astrophys. J. Lett. 812, L6

\bibitem[Pogorelov et al.(2015)]{Pogo15}
Pogorelov, N.~V., Borovikov, S.~N., Heerikhuisen, J., Zhang, M.~2015,
\apj, 812, L6

\bibitem[Pogorelov(2016)]{pogorelov_2016}
Pogorelov, N.~V. 2016, J. Phys. Conf. Ser., 719, 012013

\bibitem[Potgieter(2013)]{potgieter_2013}
Potgieter, M. 2013, Living Rev. in Solar Phys. 10, 3

\bibitem[Pohl \& Eichler(2013)]{pohl_2013}
Pohl, M. \& Eichler, D. 2013, Astrophys. J. 766, 9

\bibitem[Press et al.(1986)]{press_1986}
Press, W.H., Flannery, B.P., \& Teukolsky, S.A.\ 1986, Cambridge: University Press, 1986

\bibitem[Ptuskin(2012)]{ptuskin_2012}
Ptuskin V. 2012, Astropart. Phys. 39, 44

\bibitem[Rechester \& Rosenbluth(1978)]{rr_1978}
Rechester, A.B. \& Rosenbluth, M.N. 1978, Phys Rev Lett 40, 38

\bibitem[Rettig \& Pohl(2015)]{rettig_pohl_2015}
Rettig, R. \& Pohl, M. 2015, in Proc. of 34th ICRC, The Hague, The Netherland

\bibitem[Roberts(1956)]{Roberts}
Roberts, P.~H.~1956,
\apj, 124, 430

\bibitem[Ruderman \& Belov(2010)]{Ruderman10}
Ruderman, M.~S., \& Belov, N.~A.~2010,
J. Phys. Conf. Ser., 216, 012016

\bibitem[Ruderman \& Fahr(1995)]{Ruderman}
Ruderman, M. S., Fahr, H. J.~1995,
\aap, 299, 258

\bibitem[Santander et al.(2013)]{santander_2013}
Santander, M. et al. 2013, in Proc. of 33rd ICRC, Rio de Janeiro, Brazil; arXiv:1309.7006

\bibitem[Savchenko et al.(2015)]{savchenko_2015}
Savchenko, V., Kachelrie{\ss}, M. \& Semikoz, D.V. 2015, DOI:10.1088/2041-8205/809/2/L23

\bibitem[Scherer et al.(2016)]{scherer_2016}
Scherer K. et al. 2016, Astrop. Phys. 82, 93.

\bibitem[Shuwang et al.(2011)]{shuwang_2011}
Shuwang, C. et al. 2011, Proc. 32nd ICRC, Beijing China

\bibitem[Shaikh \& Zank(2010)]{shaikh_zank_2010}
{Shaikh}, D., \& {Zank}, G.~P. 2010, Phys. Lett. A, 374, 4538

\bibitem[Shalchi(2009)]{shalchi_2009}
Shalchi, A. 2009 Nonlinear Cosmic Ray Diffusion Theories, ASSL 362 (Springer)

\bibitem[Schwadron et al.(2014)]{schwadron_2014}
Schwadron, N.A., Adams, F.C., Christian, E.R., Desiati, P., Frisch, P., Funsten, H.O., Jokipii, J.R., McComas, D.J., Moebius, E., Zank, G.P. 2014, Science 343, 988

\bibitem[Sridhar \& Goldreich(1994)]{sg_1994}
Sridhar, S., \& Goldreich, P. 1994, Astrophys. J. 432, 612

\bibitem[Sveshnikova et al.(2013)]{sveshnikova_2013}
Sveshnikova, L.G. et al. 2013, Astropart. Phys. 50, 33

\bibitem[Xu \& Yan(2013)]{xu_yan_2013}
Xu, S., \& Yan, H. 2013, Astrophys. J. 779, 140

\bibitem[Yan \& Lazarian(2002)]{yan_lazarian_2002}
Yan, H., \& Lazarian, A. 2002, PRL, 89, 281102

\bibitem[Yan \& Lazarian(2008)]{yan_lazarian_2008}
Yan, H., \& Lazarian, A. 2008, Astrophys. J. 673, 942

\bibitem[Yan \& Lazarian(2011)]{yan_lazarian_2011}
Yan, H., \& Lazarian, A. 2011, Astrophys. J. 731, Issue 1, article id. 35, 10 pp. (2011)

\bibitem[Yu(1974)]{Yu}
Yu, G.~1974,
\apj, 194, 187

\bibitem[Zhang et al.(2009)]{zhang_2009}
Zhang, J.L. et al. 2009, Proc. 31st ICRC, \L\'od\'z, Poland

\bibitem[Zhang et al.(2014)]{zhang_2014}
Zhang, M., Zuo, P., \& Pogorelov, N. 2014, Astrophys. J. 790, 5

\bibitem[Zank et al.(1996)]{zank96}
Zank, G.~P. et al. 1996, J. of Geophys. Res., 101, 21639

\bibitem[Zank(1999)]{Zank99}
Zank, G. P. 1999, in AIP Conf. Ser. 471, Solar Wind 9, ed. S. R. Habbal et al.
(Melville, NY: AIP), 783

\bibitem[Zank et al. (2009)]{zank09}
Zank, G.~P. et al. 2009, Sp. Sci. Rev. 146, 295

\bibitem[Zank et al.(2013)]{Zank13}
Zank, G. P., Heerikhuisen, J., Wood, B. E., et al.~2013,
\apj, 763, 20




\end{thebibliography}
\end{document}